\documentclass[AMA,STIX1COL]{WileyNJD-v2}
\usepackage[utf8]{inputenc}
\usepackage{graphicx}
\usepackage{color}
\usepackage{amssymb}

\newcommand\BibTeX{{\rmfamily B\kern-.05em \textsc{i\kern-.025em b}\kern-.08em
T\kern-.1667em\lower.7ex\hbox{E}\kern-.125emX}}

\received{<day> <Month>, <year>}
\revised{<day> <Month>, <year>}
\accepted{<day> <Month>, <year>}

\begin{document}

\title{\noindent Accounting for Time Dependency in Meta-Analyses of Concordance Probability Estimates}

\author[1]{Matthias Schmid*}

\author[2]{Tim Friede}

\author[3]{Nadja Klein}

\author[1]{Leonie Weinhold}

\authormark{Matthias Schmid \textsc{et al}}

\address[1]{\orgdiv{Department of Medical Biometry, Informatics, and Epidemiology}, \orgname{University Hospital Bonn}, \orgaddress{\state{Bonn}, \country{Germany}}}

\address[2]{\orgdiv{Department of Medical Statistics}, \orgname{University Medical Center Göttingen}, \orgaddress{\state{Göttingen}, \country{Germany}}}

\address[3]{\orgdiv{Chair of Statistics and Data Science}, \orgname{Humboldt-Universität zu Berlin}, \orgaddress{\state{Berlin}, \country{Germany}}}

\corres{*Matthias Schmid, Department of Medical Biometry, Informatics, and Epidemiology, University Hospital Bonn, Venusberg-Campus 1, 53127 Bonn, Germany. \email{matthias.c.schmid@uni-bonn.de}}

%\presentaddress{Present address}

\abstract[Abstract]{Recent years have seen the development of many novel scoring tools for disease prognosis and prediction. To become accepted for use in clinical applications, these tools have to be validated on external data. In practice,  validation is often hampered by logistical issues, resulting in multiple small-sized validation studies. It is therefore necessary to synthesize the results of these studies using techniques for meta-analysis. Here we consider strategies for meta-analyzing the concordance probability for time-to-event data (``$C$-index''), which has become a popular tool to evaluate the discriminatory power of prediction models with a right-censored outcome. We show that standard meta-analysis of the $C$-index may lead to biased results, as the magnitude of the concordance probability depends on the length of the time interval used for evaluation (defined e.g.\@ by the follow-up time, which might differ considerably between studies). To address this issue, we propose a set of methods for random-effects meta-regression that incorporate time directly as covariate in the model equation. In addition to analyzing nonlinear time trends via fractional polynomial, spline, and exponential decay models, we provide recommendations on suitable transformations of the $C$-index before meta-regression. Our results suggest that the $C$-index is best meta-analyzed using fractional polynomial meta-regression with logit-transformed $C$-index values. Classical random-effects meta-analysis (not considering time as covariate) is demonstrated to be a suitable alternative when follow-up times are small. Our findings have implications for the reporting of $C$-index values in future studies, which should include information on the length of the time interval underlying the calculations.}

\keywords{Concordance probability; Fractional polynomials; Meta-regression; Prognostic factor research; Restricted cubic splines; Time-to-event data}

\jnlcitation{\cname{%
\author{M. Schmid}, 
\author{T. Friede}, 
\author{N. Klein},  and 
\author{L. Weinhold}} (\cyear{XXXX}), 
\ctitle{Accounting for Time Dependency in Meta-Analyses of Concordance Probability Estimates}, \cjournal{XXXX}, \cvol{XXXX}.}

\maketitle

%\footnotetext{\textbf{Abbreviations:} ANA, anti-nuclear antibodies; APC, antigen-presenting cells; IRF, interferon regulatory factor}

\section{Introduction}
\label{sec:1}

During the past decades, the volume of published research has increased dramatically \citep{fireGuestrin}. Even before the COVID-19 pandemic, the number of research articles has been estimated to grow by $8$-$9\%$ each year, including more than 1 million papers per year in the biomedical field alone \citep{landhuis}. At the same time, hundreds of newly ranked journals have appeared, with the estimated total amount of active peer-reviewed journals exceeding 30,000 \citep{fireGuestrin, altbachDeWit}. In view of this ``information overload'' \citep{landhuis}, there is an obvious need for evidence synthesis to ``clarify what is known from research evidence to inform policy, practice and personal decision making and improved methods for meta-analysis'' \citep{gough}.

{\it Prognostic factor research} \citep{rileyPrognostic} is a rapidly evolving field with an increased need for meta-analysis. In this field, studies aim at analyzing the associations of one or several factors (often termed ``risk factors'') with a time-to-event outcome $T\in \mathbb{R}^+$. In medicine and epidemiology, for instance, prognostic factors are often given by patient characteristics (e.g.\@ age, sex, smoking behavior, blood pressure) collected at the baseline examination of a longitudinal study. These variables might then be used to predict the occurrence of events such as death, tumor progression, or adverse events. Often, several prognostic factors are summarized by a multivariable risk score   (defined, e.g., by a linear combination of the factors).
Popular examples of risk scores are the European System for Cardiac Operative Risk Evaluation (EuroSCORE) II to predict mortality after cardiac surgery and the Framingham Risk Score for predicting coronary heart disease \citep{debrayFramework}. Score development is usually based on a statistical modeling technique applied to a set of training data, yielding a prediction model that is defined by a (univariable or multivariable) prognostic score $\eta \in \mathbb{R}$.

A key issue for the acceptability of a prognostic score is its repeated validation on externally collected test data \citep{collinsTripod, Steyerberg2019}. These validation steps have become a gold standard in prognostic modeling, as they provide a much more realistic assessment of the score's performance than would have been possible using the training data only. Importantly, the results of external validation steps are often found to be heterogeneous, showing a high variability in prognostic performance. Validation studies involving external test data might, for instance, be affected by small sample sizes and differences in the characteristics of the patient population compared to the training data \citep{debrayFramework, debrayGuide}. As a consequence, systematic reviews and meta-analyses are   ``urgently needed to summarize [the] evidence [of prediction models] and to better understand under what circumstances developed models perform adequately or require further adjustments'' \citep{debrayFramework}.

In this paper we consider strategies for meta-analyzing the {\it concordance probability for time-to-event data} (``$C$-index''), which is a widely used measure to evaluate prediction models with a time-to-event outcome \citep{harrell1984, heagertyZheng, gerds2013}. The $C$-index is a discrimination measure that compares the rankings of the individual score values $\eta_i$ and the event times $T_i$, $i=1,\ldots, n$, in a test sample of size~$n$. It is defined \citep{gerds2013} by 
\begin{equation} \label{cindex}
C (\tau) = \mbox{P}(\eta_i > \eta_j \, | \, T_i < T_j \, , \, T_i \le \tau ) \, ,
\end{equation}
where $i,j$ denote two independent observations in the test data and $\tau > 0$ is a truncation time (e.g.\@ the maximum follow-up time of a clinical study). Setting $\tau= \infty$ yields the {\it unrestricted concordance probability} $\mbox{P}(\eta_i > \eta_j \, | \, T_i < T_j)$. Generally, $C (\infty)$ takes the value 1 if the rankings of $-\eta_i$ and $T_i$ agree perfectly. Conversely, $C (\infty) = 0.5$ if $\eta$ does not predict better than chance alone. In the absence of censoring, the concordance probability can readily be evaluated by comparing all pairs $T_i , T_j$ and by estimating the conditional probability in \eqref{cindex} by its respective relative frequency in the test data. If censoring is present, however, a comparison of all pairs $T_i , T_j$ is no longer possible, and estimation of the concordance probability requires additional assumptions on the data-generating process \cite{heagertyZheng, gonenHeller, uno, gerds2013}, also see Schmid and Potapov\cite{schmidPotapov} for a comparison of estimators.

For meta-analysis, Debray et al.\@ \cite{debrayFramework} recently introduced a framework that includes, among other techniques, a method to summarize estimates of the $C$-index obtained from multiple validation studies. Based on earlier work by Snell et al.\@ \cite{snellLogit}, the authors proposed to transform estimates to the logit scale before meta-analysis. This strategy has also been adopted in several recent systematic reviews and meta-analyses\cite{doorn2017, boorn2018, he2019}. In other studies, the $C$-index was meta-analyzed on the original (untransformed) probability scale\cite{buttner2021, kothari2021}. Meta-analysis of the $C$-index using individual participant data has been studied by Pennells et al.\cite{pennells}. Hattori and Zhou \cite{hattoriZhou} proposed to construct a synthesized $C$-index from an estimate of the summary cumulative ROC curve obtained by analyzing study-specific Kaplan-Meier curves.

\begin{figure}
\centering
\includegraphics[width=12cm]{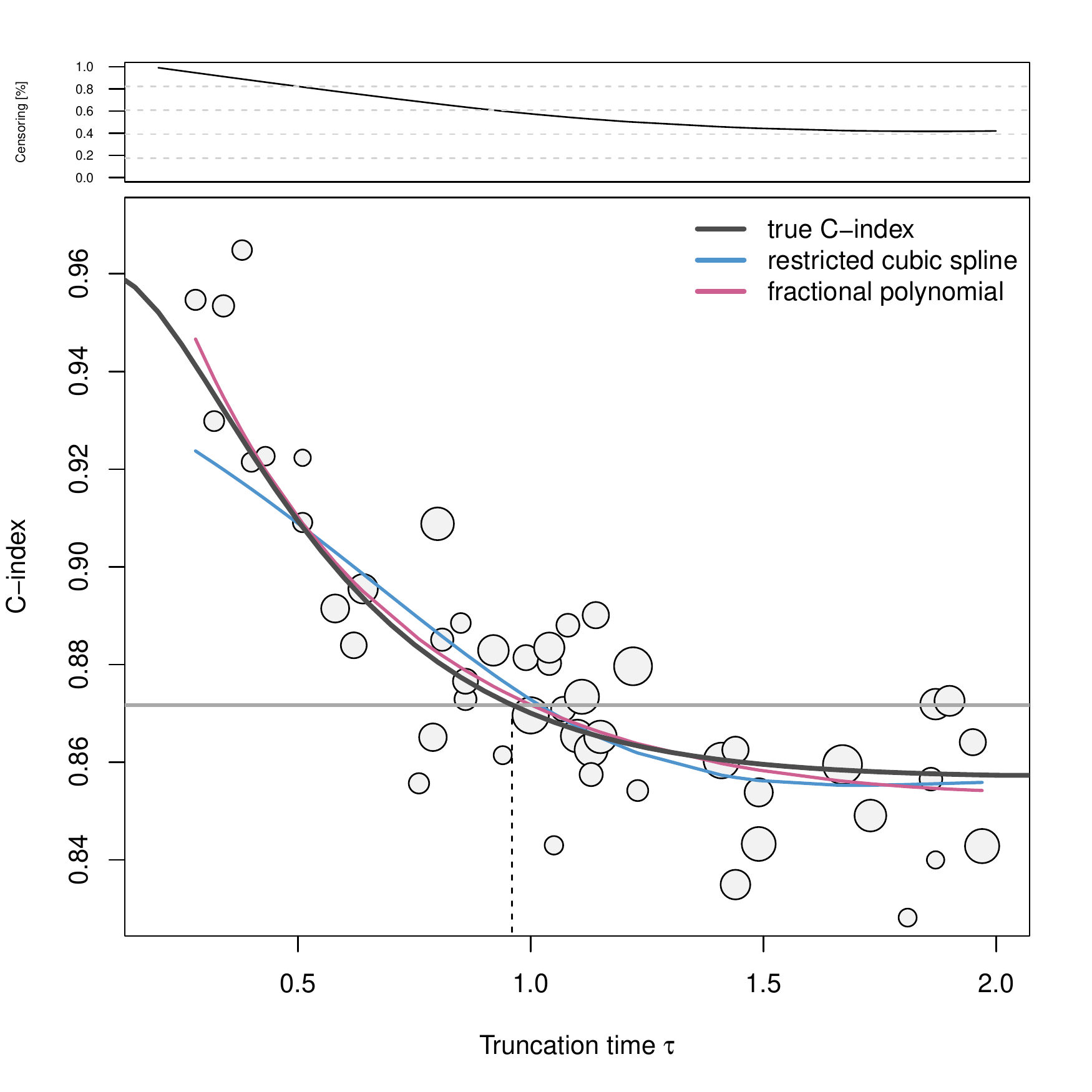}
\caption[Figure1]{
Exemplary meta-analysis of 30 studies with simulated test data. For each study, we generated event times from a Weibull accelerated failure time model of the form $\log (T) =  X -  \epsilon$, where $X$ was a normally distributed covariate with zero mean and standard deviation $0.5$, and $\epsilon$ followed a standard Gumbel distribution. Censoring times were independent of $T$ and followed an exponential distribution with rate 0.5. 
Sample sizes of the studies were generated randomly and ranged between $100$ and $1000$. After data generation, the observed event times were truncated at study-specific truncation (= maximum follow-up) times $\tau_k$, $k=1,\ldots, 30$, which were sampled from a uniform distribution on $[0.1, 2]$. The upper panel shows the expected censoring rate at each value of the truncation time. The lower panel shows the study-specific $C$-index estimates (calculated using the estimator by Uno et al.\cite{uno}). The sizes of the bubbles are proportional to the inverse variances of the $C$-index estimates. The solid black line refers to the true $C$-index according to the data-generating process (Equation (\ref{cindex})) whereas the horizontal gray line refers to the pooled $C$-index estimate that would have been obtained from a standard random effects meta-analysis ignoring time dependency. The vertical dashed line shows the ``implicit'' truncation time corresponding to the pooled estimate. Obviously, this model lacks a well defined estimand, and it is unclear how the pooled estimate should be interpreted. The blue and red lines refer to the meta-regression curves obtained from fitting a restricted cubic spline and a fractional polynomial model to the logit-transformed $C$-index estimates. For details on model specification, see Section \ref{sec:3}.
}
\label{fig1:intro}
\end{figure}

Despite numerous methodological advances, which have led to the publication of several guidance papers \citep{debrayFramework, debrayGuide}, meta-analysis of prognostic validation studies remains a challenging task. This is, in particular, due to the fact that measures of prediction accuracy in prognostic research are often related to a specific time point or time span \citep{rileyPrognostic}. Consequently, meta-analysis of validation studies becomes intrinsically difficult when study-specific performance estimates refer to different time points or spans. As seen from \eqref{cindex}, this time dependency also affects the $C$-index studied in this paper:
Since the magnitude of $C$ depends on the truncation time $\tau$, $C$-index estimates may not be comparable across studies if they relate to different values of $\tau$. Specifically, since the value of~$\tau$ is often determined by the duration of the study generating the test data, different study durations may implicitly lead to systematic differences between the resulting $C$-index estimates. Consider, for instance, the simulated meta-analysis shown in Figure \ref{fig1:intro}: In this example, $C( \tau )$ is seen to decrease with $\tau$, and the pooled estimate obtained from standard meta-analysis relates to an implicitly defined truncation time. Thus, if not accounted for, the time dependency of $C(\tau )$ may compromise both the specification of a properly defined estimand {\it and} the validity of the pooled estimate.

To address these issues and to improve the interpretability of pooled $C$-index estimates, we consider a set of statistical techniques that incorporate the time dependency of $C (\tau )$ directly in a suitably specified meta-regression model. Our proposed model is based on the frequentist modeling approach with restricted maximum likelihood (REML) estimation, as recommended in the recent guidance paper by Debray et al.\cite{debrayFramework}. Specifically, due to the above-mentioned heterogeneity of external validation results, we will focus throughout on random-effects models.
We propose to model the time dependency of $C( \tau )$ by either a restricted cubic spline (RCS) or a 2nd degree fractional polynomial (FP2), thereby accounting for nonlinearities in the regression curve (cf.\@ Figure~\ref{fig1:intro}). Using simulation studies, we will compare the RCS and FP2 models to standard random-effects meta-analysis not including $\tau$ as covariate, and also to linear meta-regression. Furthermore, we will investigate whether meta-regression can be improved by transforming the $C$-index estimates before model fitting (for instance, using a logit transformation).

The rest of the paper is organized as follows: After starting with the definition of relevant quantities (Section \ref{sec:21}), we provide a brief overview of existing techniques to estimate the $C$-index (Section \ref{sec:22}). The proposed methodology is described in Sections~\ref{sec:23} and \ref{sec:24}. Section \ref{sec:3} contains a comprehensive simulation study on the properties of the proposed approach, including a comparison to existing methods. A real-world illustration on data collected for the German Chronic Kidney Disease Study \citep{eckardt} is presented in Section \ref{sec:4}. The final section summarizes the main findings of the article.

\section{Methods}
\label{sec:2}

\subsection{Derivation and properties of the $C$-index}
\label{sec:21}

Consider a validation study with $n$ observations and a time-to-event outcome that might be subject to right censoring. The observations are assumed to be independent and identically distributed. The score values and observed event times are denoted by $\eta_i$ and $\tilde{T}_i = \min (T_i, C_i )$, $i=1,\ldots , n$, respectively, where $(C_1,\ldots , C_n)$ is a vector of continuous censoring times. The binary variables $\Delta_i = \mbox{I}(T_i \le C_i)$, $i=1,\ldots , n$, indicate whether observations are censored ($\Delta_i = 0$) or not ($\Delta_i = 1$). Assumptions on the censoring process are given below. We further assume that there are no tied observations, i.e.\@ all sample values of $T_i$ and $C_i$ are assumed to be unique.% independent of $T_1,\ldots , T_n$.

As shown by Heagerty and Zheng\cite{heagertyZheng}, the concordance probability in~(\ref{cindex}) can be derived from a set of time-dependent sensitivities and specificities, which, at each time point $t$, relate the current survival status to the event that $\eta$ exceeds a given threshold $c \in \mathbb{R}$. More specifically, following the {\it incident/dynamic} approach \citep{heagertyZheng}, one defines incident cases by observations experiencing an event at $t$ (i.e., $T_i =t$) and dynamic controls by observations having the event after $t$ (i.e., $T_i >t$). With these definitions, time-dependent sensitivities and specificities are given by
\begin{eqnarray}
\mbox{sens}_t^I (c) &=& \mbox{P}(\eta_i > c \, | \, T_i = t) \ \  \mbox{and}\\
\mbox{spec}_t^D (c) &=& \mbox{P}(\eta_i \le c \, | \, T_i > t) \, ,
\end{eqnarray}
respectively. At each time point, $\mbox{sens}_t^I (c)$ and $\mbox{spec}_t^D (c)$ can be summarized by an incident/dynamic receiver operating characteristic (ROC) curve, which is defined as
\begin{equation} \label{IDroc}
\mbox{ROC}_t^{I/D}(p) = \mbox{sens}_t^I [ (1-\mbox{spec}_t^D )^{-1}(p)] \, , \ \  p\in[0,1] \, .
\end{equation}
Incident/dynamic  ROC curves can further be summarized by the incident/dynamic AUC curve
\begin{equation} \label{IDauc}
\mbox{AUC}_t^{I/D} = \int_0^1 \mbox{ROC}_t^{I/D}(p) \, dp \, ,
\end{equation}
which equals the probability $\mbox{P}(\eta_i > \eta_j \, | \, T_i=t \, , \, T_j > t)$ for independent observations $i$ and $j$. Finally, denoting the probability density function of $T$ by $f(t)$, the concordance probability $C(\tau)$ is derived as the area under a weighted version of the incident/dynamic AUC curve. More specifically, it can be shown that %$C(\tau )$ is given by
\begin{equation} \label{derivationC}
C(\tau ) = \mbox{P}(\eta_i > \eta_j \, | \, T_i < T_j \, , \, T_i \le \tau ) = \int_0^\tau w_t^\tau \cdot \mbox{AUC}_t^{I/D} \, dt
\end{equation}
with weights $w_t^\tau = f(t)\cdot \mbox{P}(T>t) / \int_0^\tau f(u)\cdot \mbox{P} (T>u)\, du$ (see Heagerty and Zheng\cite{heagertyZheng} for a formal proof). 

A related quantity is the {\it cumulative/dynamic} ROC curve, which is defined in the same way as (\ref{IDroc}) but with $\mbox{sens}_t^I (c)$ replaced by time-dependent sensitivities of the form $\mbox{sens}_t^C (c) = \mbox{P}(\eta_i > c \, | \, T_i \le t)$. With this approach,  {\it cumulative cases} are defined by observations experiencing an event at or before $t$ (i.e., $T_i \le t$). Correspondingly, the cumulative/dynamic AUC curve is given by the areas under the cumulative/dynamic ROC curves, i.e.\@ by $\mbox{AUC}_t^{C/D} = \mbox{P}(\eta_i > \eta_j \, | \, T_i \le t \, , \, T_j > t)$. Defining a generalized version of $\mbox{AUC}_t^{C/D}$ by $\mbox{AUC}_{s,t}^{C/D} = \mbox{P}(\eta_i > \eta_j \, | \, T_i\le s \, , \, T_j > t)$, it can further be shown \citep{hattoriZhou} that
\begin{equation} \label{CDrel}
\mbox{AUC}_t^{I/D} = \frac{\partial}{\partial s} \left. \mbox{AUC}_{s,t}^{C/D} \,\right|_{s=t} \cdot \frac{\mbox{P}(T\le t)}{f(t)} \, + \, \mbox{AUC}_{t}^{C/D} \, .
\end{equation}
Thus, combining equations (\ref{derivationC}) and (\ref{CDrel}), the $C$-index can be derived using either incident or cumulative case definitions.

\begin{figure}
\centering
\includegraphics[width=12cm]{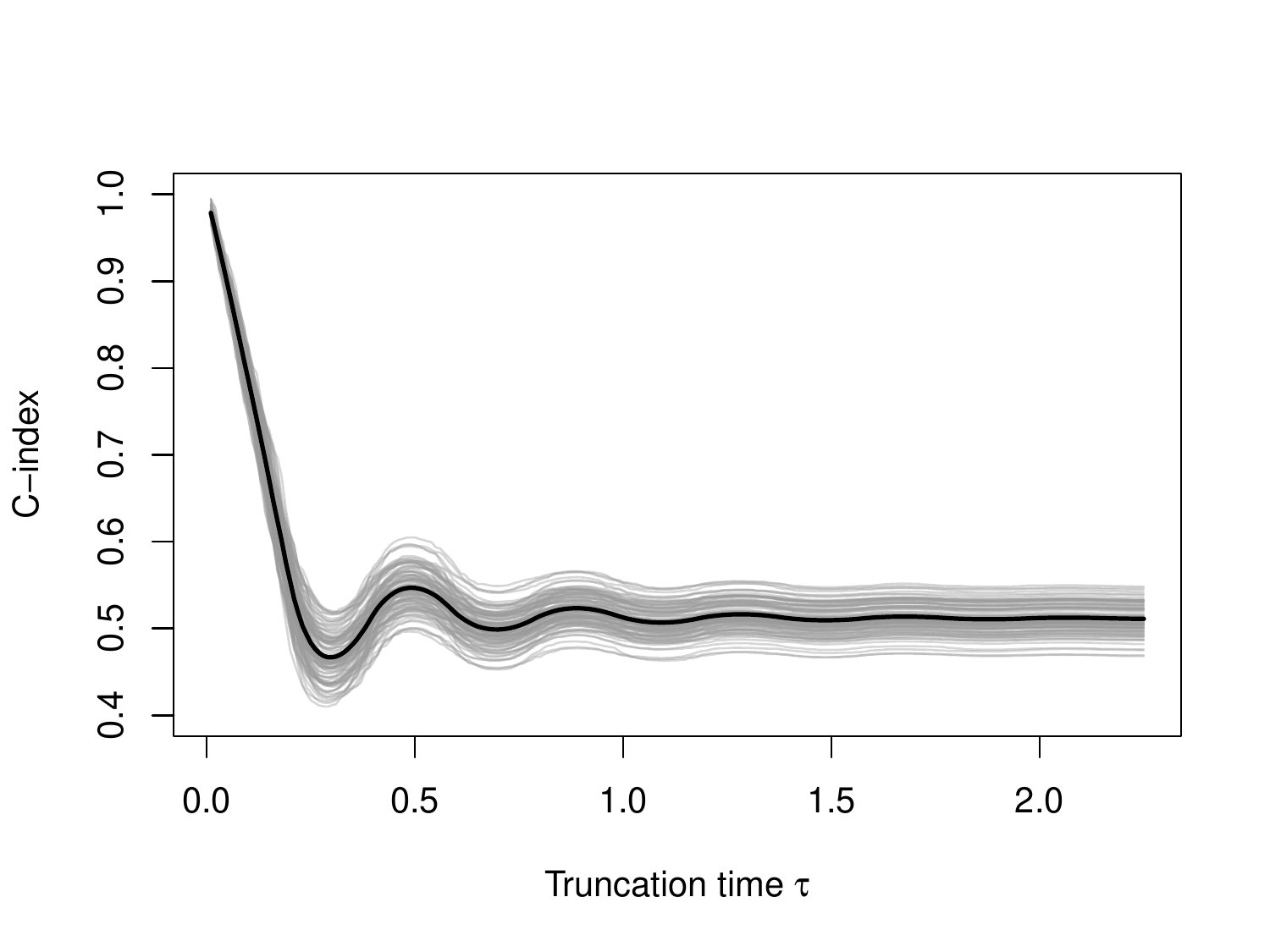}
\caption[Figure1]{
Example of a concordance probability with non-monotonic behavior. The black line (depicting the $C$-index as a function $\tau$) was derived by averaging 100 estimates of $C (\tau)$ using the method of Uno et al.\cite{uno}. Estimates were obtained from 100 independent samples with exponentially distributed event times ($n=1000$, rate = 1, no censoring). The true underlying model was given by $\eta = - \sin (8\cdot T)^2$. The gray lines refer to the 100 sample-specific curves. Although this example has been designed for illustrative purposes only and would rarely be encountered in practice, it shows that monotonicity of $C( \tau )$ cannot be assumed in general.
}
\label{fig1:intro2}
\end{figure}

In practice, $C(\tau )$ is often observed to decrease monotonically with $\tau$ (e.g.\@ Figure \ref{fig1:intro}). This behavior could, for example, be caused by a monotonically decreasing AUC curve, which tends to take smaller values as $t$ increases \citep{brentnall}. Note, however, that the monotonicity of $C(\tau )$ does not hold in general and that it is possible to construct scenarios where $C(\tau )$ shows a distinctly non-monotonic behavior (see Figure \ref{fig1:intro2}).

\subsection{Estimation of the $C$-index}
\label{sec:22}

In the absence of censoring, $C(\tau )$ is naturally estimated by the relative frequency
\begin{equation}
\hat{C}_{\mathrm{RF}} (\tau ) = \frac{\sum_{i \ne j} \mbox{I}( \eta_i > \eta_j) \cdot \mbox{I}( \tilde{T}_i < \tilde{T}_j) \cdot
\mbox{I}( \tilde{T}_i \le \tau ) }{\sum_{i \ne j} \mbox{I}( \tilde{T}_i < \tilde{T}_j) \cdot
\mbox{I}( \tilde{T}_i \le \tau )} \, ,
\end{equation}
which compares the orderings of $\tilde{T}_i$ and $\eta_i$ in an observation-wise manner. When applied to right-censored data, this approach is no longer appropriate, as pairs of observations where the shorter observed event time is censored ($\tilde{T}_i < \tilde{T}_j$ and $\Delta_i = 0$) cannot be compared in a meaningful way. An obvious way to incorporate censoring is to discard all pairs of non-comparable observations, yielding the estimator
\begin{equation}
\hat{C}_{\mathrm{Harrell}} (\tau ) = \frac{\sum_{i \ne j} \mbox{I}( \eta_i > \eta_j) \cdot \mbox{I}( \tilde{T}_i < \tilde{T}_j) \cdot
\mbox{I}( \tilde{T}_i \le \tau ) \cdot \Delta_i }{\sum_{i \ne j} \mbox{I}( \tilde{T}_i < \tilde{T}_j) \cdot
\mbox{I}( \tilde{T}_i \le \tau ) \cdot \Delta_i} \, .
\end{equation}
During the past decades, this estimator (also termed ``Harrell's $C$'') has become the most popular way to evaluate $C( \tau) $. However, it shows a notable upward bias if censoring rates are high \citep{gerds2013, schmidPotapov}. To address this issue, Uno et al. proposed an inverse-probability-of-censoring-weighted version of Harrell's $C$ (termed ``Uno's $C$'') that is defined by
\begin{equation} \label{unoC}
\hat{C}_{\mathrm{Uno}} (\tau ) = \frac{\sum_{i \ne j} \mbox{I}( \eta_i > \eta_j) \cdot \mbox{I}( \tilde{T}_i < \tilde{T}_j) \cdot
\mbox{I}( \tilde{T}_i \le \tau ) \cdot \Delta_i / \hat{G} (\tilde{T}_i)^2}{\sum_{i \ne j} \mbox{I}( \tilde{T}_i < \tilde{T}_j) \cdot
\mbox{I}( \tilde{T}_i \le \tau ) \cdot \Delta_i / \hat{G} (\tilde{T}_i)^2} \, ,
\end{equation}
where $\hat{G}(\cdot )$ is a consistent estimator of the censoring survival function $G(t) = \mbox{P}(C_i > t)$ obtained from the validation data\cite{uno}. Usually, $G(\cdot )$ is estimated by the Kaplan-Meier method, although more complex models (e.g.\@ depending on a set of covariates) might be considered. Assuming conditionally independent censoring (i.e.\@ independence of $T_i$ and $C_i$ $\forall i$ given the covariates) and a correctly specified censoring model with ${G}(t ) > \delta > 0 \ \forall t$, Uno et al. \cite{uno} showed that $\hat{C}_{\mathrm{Uno}} (\tau )$ is weakly consistent for $C(\tau )$ as $n\to\infty$. \\

\noindent {\it Remark:} The estimator considered by Gerds et al. is slightly different from (\ref{unoC}) in that $\hat{G} (\tilde{T}_i)^2$ is replaced by $\hat{G} (\tilde{T}_i) \cdot \hat{G} (\tilde{T}_i-)$ in both the numerator and the denominator, where $\tilde{T}_i-$ refers to a time point that is infinitesimally smaller than $\tilde{T}_i$\cite{gerds2013}. Clearly, this difference is only relevant when $\hat{G}(\cdot )$ is not continuous in $t$ (for instance when the Kaplan-Meier method is used to estimate~$\hat{G}(\cdot )$). In our analysis we will use the R add-on package {\bf pec} \citep{pec} that implements the method by Gerds et al.\cite{gerds2013} but refer to this estimator as ``Uno's $C$''.\\ 

A major advantage of Harrell's $C$ and Uno's $C$ is that both estimators are non-parametric in the sense that they do not make any assumptions on the distribution of $T$. An alternative way to deal with non-comparable pairs of observations is to specify a parametric or semi-parametric working model for $T$ (e.g.\@ a Cox regression model) and to derive estimators of $C(\tau )$ based on the characteristics of this model\cite{heagertyZheng, gonenHeller, songZhou}. It is also possible to apply a model-free estimator of the incident/dynamic AUC curve \citep{vanGeloven} and to estimate the $C$-index via numerical integration of the AUC estimate. In this paper we will consider Harrell's $C$ and Uno's $C$ throughout.

\subsection{On the role of the truncation time $\tau$}
\label{sec:23}

As stated in Section \ref{sec:1}, the unrestricted $C$-index $\mbox{P}(\eta_i > \eta_j \, | \, T_i < T_j)$ comes with an intuitive probabilistic interpretation, comparing the rankings of the values $\eta_i$ and $T_i$, $i=1,\ldots , n$. This interpretation is considerably less intuitive if an additional truncation time $\tau < \infty$ is included in the definition of the $C$-index. Nonetheless, there exist both conceptual {\it and} technical reasons to prefer a restricted version of the concordance probability over the unrestricted one: First, the sample values $(\tilde{T}_i , \Delta_i , \eta_i)$, $i=1,\ldots , n$, are often obtained from a validation study with a limited follow-up time. In this case, the maximum possible time horizon~$\tau$ is naturally given by the length of the follow-up time, implying that {\it any} estimate of the concordance probability derived from the validation data is a restricted one\cite{longato}. Second, the censoring model used in the definition of Uno's $C$ usually assumes $G(t) > \delta > 0 \ \forall t$, posing a problem if non- or semi-parametric methods are applied to estimate $G(\cdot)$ beyond $\tau := \max_i (\tilde{T}_i )$. In particular, the Kaplan-Meier estimator (being the predominant estimator of $G(\cdot )$ in practice) is zero beyond $\max_i (\tilde{T}_i )$ if the longest observed event time corresponds to a censored observation (and does not even exist beyond $\max_i (\tilde{T}_i )$ if this observation has $\Delta_i = 1$). These problems can be avoided if a restricted version of the $C$-index (with a suitably defined value of $\tau < \max_i (\tilde{T}_i )$) is considered for analysis.

\subsection{Meta-regression of $C$-index estimates}
\label{sec:24}

In this section we describe a set of models to account for the time dependency of the restricted $C$-index in meta-regression. We start with the classical model for time-independent meta-analysis, also discussing possible transformations of $C$-index estimates before model fitting.\\

\noindent {\it Random-effects meta-analysis.} Consider a set of $K$ independent validation studies with study-specific estimates $\hat{C}_1, \ldots , \hat{C}_K$ and variance estimates $\hat{\sigma}_1^2 , \ldots , \hat{\sigma}_K^2$. As argued above, each of these estimates relates to a study-specific truncation time $\tau_k$, \linebreak $k=1,\ldots , K$. Classical parametric meta-analysis ignores this time dependency, assuming that $\hat{C}_1, \ldots , \hat{C}_K$ are estimates of some study-specific unrestricted concordance probabilities $C_1, \ldots , C_K$. We further assume (here and in all other models, following standard procedures) that each $\hat{\sigma}_k^2$ corresponds to the true variance $\sigma_k^2$ of the respective residual term $\epsilon_k = \hat{C}_k - C_k$. 
The corresponding model is given by
\begin{eqnarray} \label{eqn:classicMeta}
\hat{C}_k &=& C_k + \epsilon_k \, , \, \epsilon_k \sim N(0, \sigma_k^2) \, , \nonumber\\
C_k &=& C_{\text{pop}} + a_k \, , \, a_k \sim N(0, \sigma_a^2) \, , \, k = 1,\ldots , K,
\end{eqnarray}
where the aim is to obtain a ``pooled'' estimate of the population value $C_{\text{pop}}$. The study-specific deviations $a_k$ are assumed to be independent of $\epsilon_k$ and to follow a normal distribution with between-study variance $\sigma_a^2$. 

If homogeneity of studies is assumed, i.e.\@ $\sigma_a^2 = 0$ and $C_1 = \ldots = C_k= C_{\text{pop}}$, then this is referred to as common-effect meta-analysis. In contrast, random-effects meta-analysis assumes $\sigma_a^2 \neq 0$, accounting for study-specific heterogeneity. 
Results of validation studies are usually expected to vary between studies, as these may differ in sample selection and many other design aspects. Therefore, and in line with the recommendation of Debray et al.\cite{debrayFramework}, we will restrict our analysis to random-effects models for the purpose of our study. In the literature, numerous methods to estimate $\sigma_a^2$ have been proposed \citep{sinha2011statistical}. Here we follow the recommendation by Debray et al.\cite{debrayFramework} and consider methods based on restricted maximum likelihood (REML) estimation. With this approach, estimation of $\sigma_a^2$ and $C_{\text{pop}}$ is performed jointly using a model with inverse variance weights $1/\hat{\sigma}_k^2$.\\

\noindent \textit{Transformations of $C$-index estimates.} The classical approach to meta-analyze $C$-index values is based on the untransformed estimates $\hat{C}_1, \ldots , \hat{C}_K$. This approach, which relies on the asymptotic normality of estimators like Uno's $C$, has been followed e.g.\@ by Büttner et al.\cite{buttner2021} and Waldron et al.\cite{waldron2014}. Other authors have argued that the concordance probability is bounded between 0 and 1, so that the normality and homoscedasticity assumptions in \eqref{eqn:classicMeta} are unlikely to hold. To address these issues, they transformed $C$-index estimates before meta-analysis, using e.g.\@ the logistic transformation $g(\hat{C}_k)=\log(\hat{C}_k/(1-\hat{C}_k))$ \citep{vanKlaveren2014} or the arcsine square root transformation $g(\hat{C}_k)=\sin^{-1}(\hat{C}_k^{1/2})$ \citep{vanKlaveren2014, schwarzer2019}. After model fitting, the estimate of $C_{\text{pop}}$ is usually back-transformed to the original probability scale.\\

\noindent \textit{Linear meta-regression.} As argued above, classical meta-analysis does not account for the implicit time dependency of the estimates $\hat{C}_1 , \ldots , \hat{C}_K$. As a consequence, it is unclear how to interpret the population value $C_{\text{pop}}$ in Equation (\ref{eqn:classicMeta}). In particular, $C_{\text{pop}}$ will not be a meaningful approximation of the unrestricted $C$-index if $C(\tau )$ decreases with $\tau$ (see Figure \ref{fig1:intro}).

A more appropriate approach to account for the time dependency of $C(\tau ) $ is to consider a meta-regression model of the form
\begin{eqnarray} \label{eqn:classicMetaReg}
g(\hat{C}_k) = f(\tau_k; \gamma) + a_k + \epsilon_k \, , \ \ a_k~\sim~N(0, \sigma_a^2) \, , \ \ \epsilon_k~\sim~N(0, \sigma_k^2) \, ,
\end{eqnarray}
$k = 1,\ldots , K$, where $g(\cdot )$ is a pre-specified transformation (for instance, the logistic transformation) and $\tau_k$ is included as a covariate. The relationship between $\hat{C}_k$ and $\tau_k$ is modeled by the (possibly nonlinear) function $f(\cdot ) $ depending on a coefficient vector $\gamma \in \mathbb{R}^p$. Instead of calculating a one-dimensional pooled estimate of $C_{\text{pop}}$, the idea is to first estimate the coefficient vector $\gamma$ and to subsequently approximate the full curve $C(\tau ) $ by the estimated regression function $f(\tau; \hat{\gamma})$.

The simplest way of specifying a model of the form (\ref{eqn:classicMetaReg}) is to consider the linear function $f(\tau_k; \gamma) = \gamma_0 + \tau_k \cdot \gamma_1$, yielding the {\it linear meta-regression model} with $\gamma = (\gamma_0 , \gamma_1 )^\top \in \mathbb{R}^2$. Estimation of $\gamma$ is performed in the same way as above, i.e.\@ using REML with inverse variance weights $1/\hat{\sigma}_k^2$.\\

% truncated power basis function aus rcs man page: rcspline.eval
\noindent {\it Spline meta-regression.} Although the linear meta-regression model accounts for the time dependency of $C (\tau )$, it does not capture nonlinear functional relationships as the ones presented in Figures \ref{fig1:intro} and \ref{fig1:intro2}. This might be a problem even when the values of $\hat{C}_k$ are transformed before model fitting. A convenient approach to address nonlinearity is to represent $f(\tau_k ; \gamma)$ by a restricted cubic spline, as implemented in the R packages \textbf{metafor} and \textbf{rms} \citep{metafor, rms}. With this approach, $ f(\tau_k; \gamma)$ is specified as a weighted sum of truncated power basis functions (defined using a pre-specified set of {\it interior knots}), and $\gamma$ is set equal to the vector of weights. Regarding the number and placement of the knots, we follow the recommendations in Section~2.4.6 of Harrell Jr. \citep{harrell2001regression}, using four knots (= basis functions) if $K \ge 30$ and three knots if $K < 30$. Obviously, the spline meta-regression model depends on a larger number of coefficients than the linear meta-regression model, increasing its flexibility but also being more prone to overfitting (especially when the number of studies is small).\\

\noindent {\it Fractional polynomial meta-regression.} 
An alternative to spline regression is fractional polynomial (FP) modeling, which is based on transformations of $\tau_k$ by a weighted sum of power functions. Following  Royston and Sauerbrei\cite{mfpbook}, we consider fractional polynomials of degree 2 (``FP2''), which are defined by $f(\tau_k ; \gamma ) = \gamma_0 + \gamma_1 \cdot \tau_k^{p_1} + \gamma_2 \cdot \tau_k^{p_2}$, where $p_1$ and $p_2$ are chosen from the predefined set of powers $S = \{-2, -1, -0.5, 0, 0.5, 1, 2, 3\}$ with $\tau_k^0 := \log(\tau_k)$. In case $p_1=p_2=:p^*$, the function~$f(\cdot)$ is defined by $f(\tau_k ; \gamma ) = \gamma_0 + \gamma_1 \cdot \tau_k^{p^*} + \gamma_2 \cdot \tau_k^{p^*} \cdot \log(\tau_k)$. As demonstrated by Royston and Sauerbrei\cite{mfpbook} and Royston and Altman\cite{royston1994mfp}, FP2 models are able to capture a wide variety of nonlinear trends. For meta-regression of the $C$-index we propose to use a function of the form $f(\tau_k ; \gamma ) = \gamma_0 + \gamma_1 \cdot \tau_k^{-0.5} + \gamma_2 \cdot \tau_k^{0.5}$, i.e.\@ $p_1$ and $p_2$ are set to $-0.5$ and $0.5$, respectively. The latter values are inspired by the ``typical'' shape of $C(\tau )$ in Figure \ref{fig1:intro} and by the fact that this shape closely resembles the respective FP2 plot in Figure 1 of Royston\cite{roystonStata}. Section \ref{sec:4} presents a detailed empirical analysis of the choice of the power values.\\ 

\noindent {\it Exponential decay meta-regression.} In addition to the aforementioned meta-regression models, we consider the {\it exponential decay meta-regression model}, which employs an alternative regression function that requires the time-restricted concordance index to be monotone decreasing with $\tau$. This approach might be suitable when there is strong evidence of a monotonic trend in $C(\tau )$ (as the one shown in Figure \ref{fig1:intro}). 
The exponential decay meta-regression model is specified as
\begin{eqnarray}
g(\hat{C}_k ) = \theta + a_k + (R_0- (\theta + a_k)) \cdot \exp{\left(-\exp(\beta) \cdot \tau_k\right)} + \epsilon_k \, , \ \ a_k \sim N(0, \sigma_a^2) \, , \ \ \epsilon_k \sim N(0, \sigma_k^2) \, , \label{eqn:exDec}
\end{eqnarray}
with parameter vector $\gamma=(\theta, \beta, R_0)^T$. By definition, $g(\hat{C}_k )$ converges to $\theta + a_k + \epsilon_k$ as $\tau_k \to \infty$, implying that the unrestricted $C$-index might be estimated by the fitted value of $\theta$. (Note that this is not possible with the linear, spline, and fractional polynomial regression approaches described above.) The value of $R_0$ corresponds to an approximation of $C(\tau_k)$ at $\tau_k=0$, and $\beta$ determines the rate of decay. Further note that the random-effects structure of the exponential decay model is slightly different from the respective structure in (\ref{eqn:classicMetaReg}), as the random effect $a_k$ enters (\ref{eqn:exDec}) in a nonlinear way.

\section{Simulation study}
\label{sec:3}

\subsection{Experimental setup}

Here we present the results of a simulation study that we conducted to analyze the properties of the models discussed in Section~\ref{sec:2}. The aims of our study were (i) to investigate the benefit of incorporating the truncation times $\tau_k$ in meta-regression models for the concordance probability, (ii) to compare the performance of the meta-regression approaches  discussed in Section \ref{sec:2} with regard to estimation accuracy and numerical stability, and (iii) to investigate the use of variable transformations before model fitting.

Our simulation study was based on a Weibull model of the form
\begin{equation} \label{mod:Weibull}
    \log(T_i) = \eta_i - \sigma W_i \, , \ \ \eta_i \sim N(0,0.5^2) \, , \ \ i = 1,\ldots , n,
\end{equation}
with normally distributed score values $\eta_i$ and noise variables $W_i$ that followed a standard Gumbel distribution (independent of $\eta_i$). The parameter $\sigma$ was set to 0.5, yielding the $C$-index curve presented in Figure S1 in the Supporting Information. For example, we obtained $C(\tau_k) = 0.79$, $0.77$ and $0.74$ for $\tau_k = 0.2$, $0.7$ and $1.5$, respectively.

Based on Model \eqref{mod:Weibull}, we considered three scenarios for meta-regression, setting the number of validation studies to $K=15$ (``small''), $30$ (``moderate''), and $50$ (``large''). For each $K$ we simulated study data sets with $n_k$ observations ($k=1,\ldots , K$), generating the sample sizes $n_k$ randomly from the grid $ \{ 100, 110, 120, \ldots , 990, 1000 \} $. The censoring times $C_i$ were sampled from an exponential distribution with rate parameter $0.5$. The truncation times $\tau_k$ of the studies were generated as follows: First, we defined a joint maximum follow-up time (denoted by $\tau_{\text{max}}$) for all studies. Afterwards we sampled the values $\tau_k$ from a truncated gamma distribution on the interval $[0.1; \tau_{\text{max}}]$. The shape and rate parameters of this distribution were set to 1.5 and~1, respectively. Subsequently, event times with $\tilde{T_i} > \tau_k $ were censored at $\tau_k$ (study-wise). We considered three values of $\tau_{\text{max}}$, namely $\tau_{\text{max}}=0.7$ (``short follow-up''), $\tau_{\text{max}}=0.9$ (``medium follow-up''), and $\tau_{\text{max}}=2$ (``long follow-up''), yielding average censoring rates of 0.92, 0.86 and 0.64, respectively. Estimates of the $C$-index were obtained using Uno's $C$, as implemented in the R package {\bf pec}. To introduce study-specific heterogeneity, we added normally distributed random numbers $a_k$ to the $C$-index estimates. These numbers were drawn from a normal distribution with zero mean and variance $\sigma_a^2$. Again we considered three scenarios, setting $\sigma_a^2 = 0$ (``no heterogeneity''), $\sigma_a^2 = 0.01^2$ (``moderate heterogeneity''), and $\sigma_a^2 = 0.03^2$ (``large heterogeneity''). The choice of these numbers was inspired by our work in Zacharias et al.\cite{zacharias}, where differences in $C$-index values varied between 0.002 and 0.06 across validation studies (the latter number corresponding to two standard deviations of our ``large heterogeneity'' setting).

%{\color{red}muss es $0.01^2$ und $0.03^2$ heissen, vgl. Caption Tabelle 1?}

In each of the $3 \times 3 \times 3 = 27$ scenarios (defined by the values of $K$, $\tau_{\text{max}}$ and $\sigma_a^2$) we set the number of Monte Carlo replications to~1000 and fitted the following models to the simulated study data: (i) meta-analysis, (ii) linear meta-regression, (iii) spline meta-regression, (iv) fractional polynomial meta-regression, and (v) exponential decay meta-regression, as described in Section~\ref{sec:2}. Model fitting was carried out using the {\tt metamean}, {\tt metareg} and {\tt rma} functions of the R packages {\bf meta} and {\bf metafor} \citep{metafor, meta}, except for the exponential decay model for which the function {\tt nlme} of the R package {\bf nlme} \citep{nlme} was used. For sensitivity analysis, we additionally carried out random-effects meta-analyses using the $30\%$ and $50\%$ of studies with largest values of $\tau_k$ only. This approach was inspired by the shape of $C(\tau )$ in Figure \ref{fig1:intro}, assuming that studies with a long follow-up time would be less affected by time dependency due to the convergence behavior of $C(\tau )$. Standard errors of the $C$-index estimates (needed to calculate the weights $1/\hat{\sigma}_k^2$) were computed using $1000$ bootstrap samples with replacement.

Transformation functions included the identity transformation (id), the logistic transformation (logit), and the arcsine square root transformation (asin). Another candidate transformation would have been the double arcsine transformation; however, we did not consider this transformation because it has recently been found unsuitable for meta-analysis purposes \citep{roeverFriede}. Comparisons of the respective $C$-index estimates were carried out at the population level (setting $\sigma_a=0$) after back-transforming the fitted values to the original scale. 

The following criteria were used to evaluate the results of the simulation study:
\begin{itemize}
\item[(i)] To investigate the numerical stability of the methods, we calculated the proportion of simulation runs in which the respective R fitting functions issued errors and/or warnings indicating convergence issues. These assessments were necessary since each of the studies entered the models with a separate random effect $a_k$ and a separate variance term $\sigma_k^2$, potentially leading to some instabilities in the REML procedure.
\item[(ii)] To investigate the estimation accuracy of the meta-regression models at a fixed time point, we evaluated the pooled estimates of the restricted concordance probability at $t = 0.8 \cdot \tau_{\text{max}}$ and compared these estimates (including their $95\%$ confidence intervals) to the respective true values of $C(0.8 \cdot \tau_{\text{max}})$.
\item[(iii)] For all methods we computed the areas enclosed by the true and the estimated $C$-index curves, using $\min_k (\tau_k)$ and $\max_k (\tau_k)$ as interval limits. All areas were divided by the interval length $(\max_k (\tau_k)-\min_k (\tau_k))$, see Figure S2 in the Supporting Information for an illustration.
\end{itemize}

\subsection{Results}

We first present the results obtained from the scenario with $K=30$ studies. The results of the other two scenarios ($K=15$, $K=50$) are presented in the Supporting Information (Figures S3 and S4).

Table \ref{tab:failure} contains a summary of the failure rates, i.e.\@ the percentages of the simulation runs in which the R fitting functions issued either an error or a warning. It is seen that fitting the exponential decay model resulted in a large number of convergence issues, with failure rates being as high as $74.3\%$ of the simulation runs. Generally, failure rates tended to decrease with the length of follow-up, which can be explained by the more pronounced curvature of the $C$-index curve in these scenarios (showing stronger support for the shape of the exponential decay function). Still, failure rates were high even in the most favorable settings. We conclude that the numerical stability of the exponential decay method is not sufficient for meta-regression of the concordance index, and we therefore did not consider this model further. The failure rates of the other methods were throughout close to zero. 

Similar results were obtained in the scenarios with $K=15$ studies and $K=50$ studies (Tables~S1 and S2, respectively, in the Supporting Information).

\begin{table}[!ht]
\caption{Results of the simulation study ($K=30$). The table summarizes the failure rates ($\%$) of the meta-regression models described in Section \ref{sec:2}. Failure rates were defined by the percentages of simulation runs in which the respective R fitting functions issued either an error or a warning (MA = meta-analysis, linear = linear meta-regression, RCS = restricted cubic spline meta-regression, FP2 = 2nd degree fractional polynomial meta-regression).}
\begin{center}
\resizebox{0.8\textwidth}{!}{
\begin{tabular}{l|rrr|rrr|rrr}
    \hline
    &\multicolumn{3}{c}{$\sigma_a=0$}&\multicolumn{3}{c}{$\sigma_a=0.01$}&\multicolumn{3}{c}{$\sigma_a=0.03$}\\
        \hline
 &short&moderate&long & short&moderate&long &short&moderate&long \\ 
\hline
MA (id) & 0.0 & 0.0 & 0.3 & 0.3 & 0.1 & 0.1 & 0.0 & 0.0 & 0.0 \\ 
  MA (id, last 50 \%) & 0.1 & 0.0 & 0.1 & 0.0 & 0.0 & 0.0 & 0.1 & 0.0 & 0.0 \\ 
  MA (id, last 30 \%) & 0.1 & 0.0 & 0.2 & 0.0 & 0.0 & 0.2 & 0.0 & 0.2 & 0.0 \\  
    \hline
  MA (logit) & 0.4 & 0.2 & 0.0 & 0.6 & 0.0 & 0.0 & 0.2 & 0.0 & 0.0 \\ 
  MA (logit, last 50 \%) & 0.4 & 0.2 & 0.0 & 0.3 & 0.0 & 0.0 & 0.1 & 0.0 & 0.0 \\ 
  MA (logit, last 30 \%) & 0.2 & 0.0 & 0.0 & 0.4 & 0.1 & 0.0 & 0.0 & 0.0 & 0.0 \\ 
    \hline
MA (asin) & 0.1 & 0.2 & 0.4 & 0.1 & 0.0 & 0.0 & 0.0 & 0.0 & 0.0 \\ 
  MA (asin, last 50 \%) & 0.1 & 0.0 & 0.0 & 0.2 & 0.1 & 0.1 & 0.1 & 0.0 & 0.0 \\ 
  MA (asin, last 30 \%) & 0.0 & 0.1 & 0.1 & 0.2 & 0.0 & 0.1 & 0.0 & 0.0 & 0.0 \\ 
    \hline
 linear (id) & 0.5 & 0.1 & 0.0 & 0.2 & 0.1 & 0.0 & 0.0 & 0.1 & 0.0 \\ 
  linear (logit) & 0.5 & 0.2 & 0.1 & 0.7 & 0.0 & 0.0 & 0.3 & 0.1 & 0.1 \\ 
  linear (asin) & 0.4 & 0.2 & 0.0 & 0.3 & 0.3 & 0.0 & 0.1 & 0.1 & 0.1 \\  
    \hline
RCS (id) & 0.3 & 0.0 & 0.2 & 0.0 & 0.0 & 0.1 & 0.0 & 0.1 & 0.0 \\ 
  RCS (logit) & 0.1 & 0.1 & 0.0 & 0.0 & 0.0 & 0.1 & 0.1 & 0.1 & 0.1 \\ 
  RCS (asin) & 0.0 & 0.1 & 0.0 & 0.0 & 0.0 & 0.1 & 0.1 & 0.1 & 0.1 \\ 
    \hline
FP2 (id) & 0.3 & 0.2 & 0.1 & 0.0 & 0.0 & 0.0 & 0.1 & 0.0 & 0.0 \\ 
  FP2 (logit) & 0.2 & 0.1 & 0.0 & 0.3 & 0.0 & 0.0 & 0.1 & 0.1 & 0.1 \\ 
  FP2 (asin) & 0.2 & 0.0 & 0.1 & 0.1 & 0.2 & 0.1 & 0.1 & 0.2 & 0.1 \\
    \hline
exponential decay (id) & 56.7 & 43.3 & 16.2 & 56.3 & 41.1 & 16.3 & 58.4 & 42.2 & 24.5 \\ 
  exponential decay (logit) & 70.8 & 61.9 & 27.1 & 72.1 & 60.2 & 22.9 & 74.3 & 68.8 & 37.6 \\ 
  exponential decay (asin) & 62.0 & 46.6 & 17.5 & 60.7 & 46.5 & 19.5 & 62.7 & 50.9 & 32.3 \\ 
    \hline
\end{tabular}}
\end{center}
\label{tab:failure}
\end{table}

\begin{figure}[ht!]
\centering
\includegraphics[width=14cm]{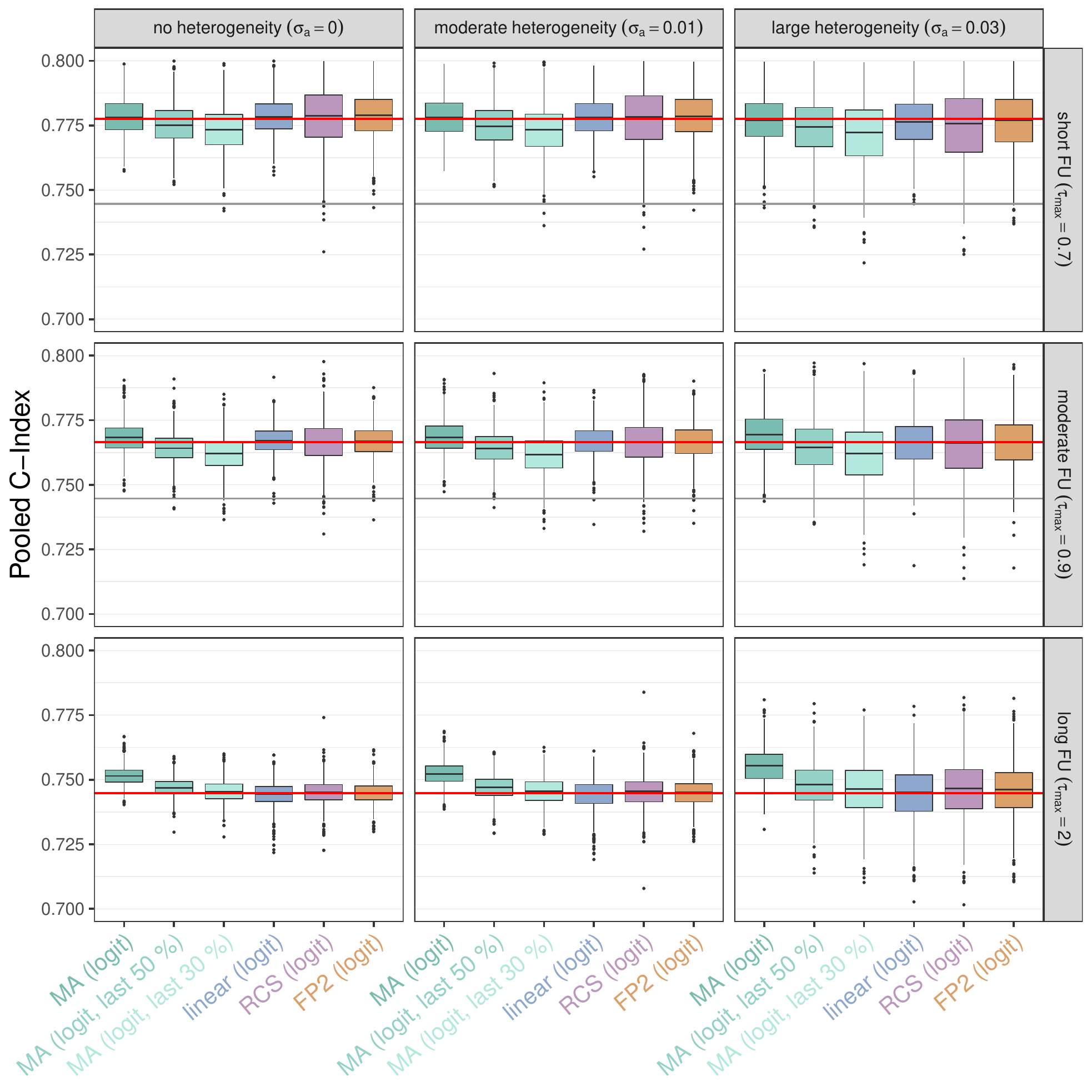}
\caption[Figure2b]{Results of the simulation study  ($K=30$). The boxplots summarize the pooled estimates of the restricted concordance index at $0.8\cdot\tau_{\text{max}}$. All $C$-index estimates were transformed using a logistic transformation before model fitting. The red and the black lines refer to the true values of $C(0.8\cdot\tau_{\text{max}})$ and the unrestricted values of the concordance index, respectively. Note that the black lines coincide with the red lines in the lower three panels.}
\label{fig2b:simresults_logit}
\end{figure}

Figure \ref{fig2b:simresults_logit} presents the pooled concordance probability estimates at the fixed truncation time $0.8\cdot\tau_{\text{max}}$ (logistic transformation, $K=30$). It is seen that ignoring the time-dependency of $C(\tau )$ resulted in a bias of classical random-effects meta-analysis. In line with Figure \ref{fig1:intro}, this bias was positive in most of the scenarios and was most pronounced when the follow-up time was long. It was close to zero on average when the follow-up time was short. As expected, the estimates obtained from the sensitivity analyses (corresponding to random-effects analyses of the $30\%$ and $50\%$ of studies with largest values of $\tau_k$) were almost unbiased in the scenarios with long follow-up. The meta-regression methods performed well in all settings, with spline meta-regression showing a higher variability than linear and fractional polynomial meta-regression. As expected, the variance of the estimates increased as the heterogeneity between studies became larger.

Similar results were obtained in the scenarios with $K=15$ and $K=50$ (Figures S3 and S4, respectively, in the Supporting Information). The results obtained from the untransformed and arcsine-square-root-transformed estimates ($K=30$) are presented in Figures S5 and S6, respectively, in the Supporting Information. Compared to the logit-transformed estimates, these estimates showed a slightly increased bias, especially in the scenarios with short follow-up. Again, spline meta-regression had a higher variability than fractional polynomial meta-regression.

\begin{figure}
\centering
\includegraphics[width=14cm]{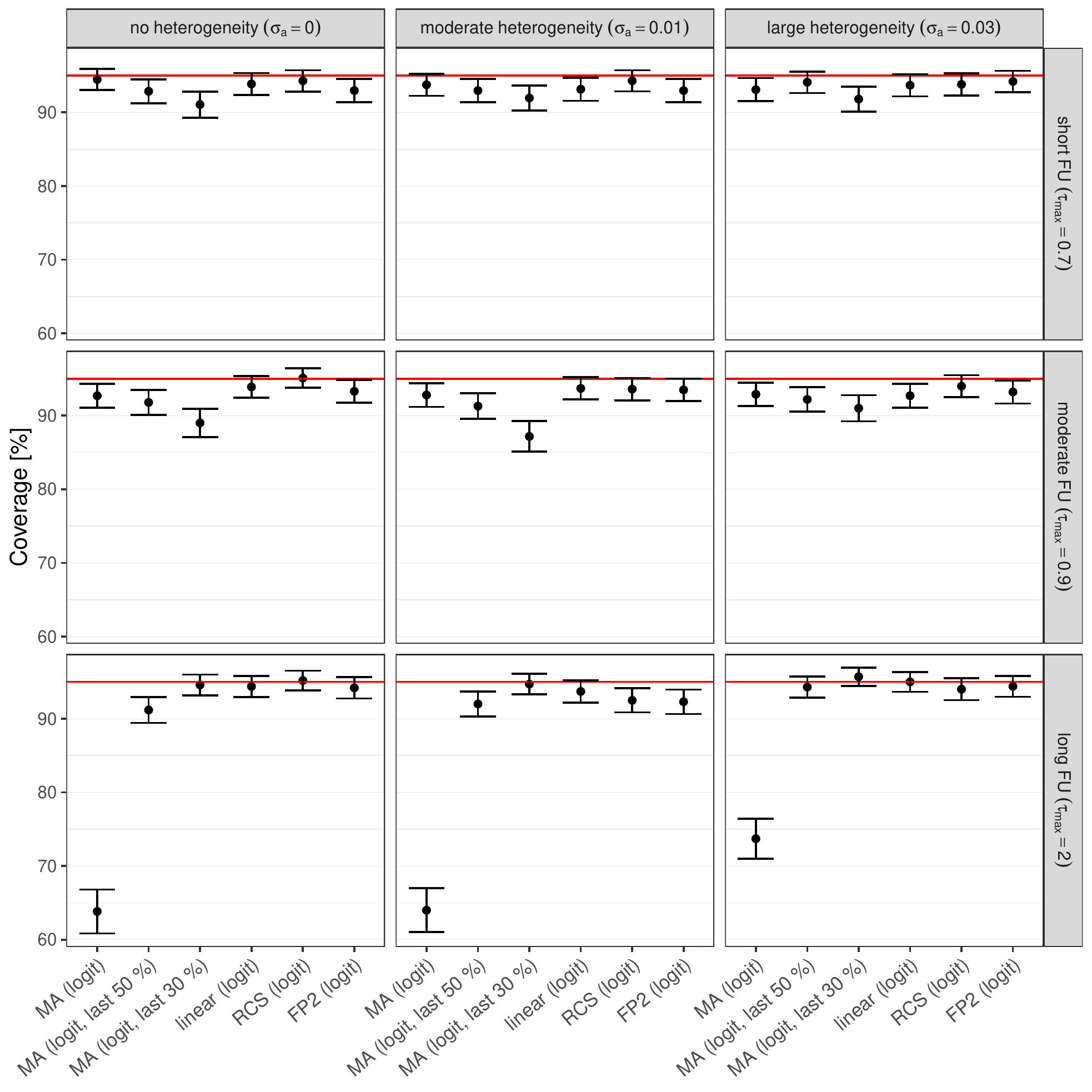}
\caption[Figure2b]{Results of the simulation study  ($K=30$). The plot shows the estimated coverage probabilities (\%), i.e.\@ the proportion of simulation runs in which the 95\% Hartung-Knapp confidence intervals contained the true value of $C(0.8\cdot\tau_{\text{max}})$. Confidence limits (represented by the black lines) were computed as $[\hat{p} \pm 1.96 \cdot \sqrt{\hat{p} \cdot (1-\hat{p}) / 1000}]$, where $\hat{p}$ denotes the point estimate of the coverage probability. The red lines refer to the $95\%$ confidence level. All $C$-index estimates were transformed by a logistic transformation before model fitting.}
\label{fig2b:coverage_logit}
\end{figure}

Figure \ref{fig2b:coverage_logit} presents the estimated coverage probabilities of the $95\%$ Hartung-Knapp confidence intervals at the fixed truncation time $0.8\cdot\tau_{\text{max}}$ (logistic transformation, $K=30$). It is seen that the confidence intervals obtained from the meta-analysis model (ignoring follow-up time) did not reach the desired coverage probability in the scenario with long follow-up. All other coverage probability estimates were close to the $95\%$ level. The results obtained from the models with untransformed and arcsine-square-root-transformed $C$-index estimates ($K=30$) showed similar patterns (Figures S7 and S8, respectively, in the Supporting Information), except that the meta-analysis model performed generally worse than the other models when fitted to the untransformed estimates (regardless of the length of follow-up).
The estimated coverage probabilities obtained from the scenarios with $K=15$ and $K=50$ showed similar patterns as well, again suggesting that the meta-analysis model is inferior to the meta-regression models in the scenarios with long follow-up (data not shown).

Table \ref{tab:area} presents the areas enclosed by the true and the estimated $C$-index curves. It is seen that the areas obtained from time-independent random-effects meta-analysis tended to increase with increasing follow-up time, whereas the respective areas obtained from the meta-regression models tended to {\it de}crease with increasing follow-up time. Random-effects meta-analysis was the overall best method in settings with short follow-up. By contrast, linear meta-regression of logit-transformed $C$-index estimates and fractional polynomial meta-regression of logit-transformed $C$-index estimates tended to perform best in the scenarios with moderate and long follow-up times, respectively. These results clearly suggest that the time-constant functions obtained from random-effects meta-analysis are reasonable approximations to $C(\tau )$ in settings with a short follow-up. By contrast, the benefits of modeling $C$-index values by a regression function become apparent when follow-up times are ``long enough'' to demonstrate possible time dependencies and nonlinear shapes of $C(\tau )$. (For a discussion on how to assess the relative length of the follow-up time, see Section \ref{sec:4}.) In most cases, the areas between the true and the estimated curves were smallest when $C$-index estimates were transformed by the logistic transformation before model fitting. Very similar results were obtained in the scenarios with $K=15$ and $K=50$ (Tables S3 and S4, respectively, in the Supporting Information).\\

\begin{table}[!t]
\caption{Results of the simulation study ($K=30$). The table summarizes the areas enclosed by the true and the estimated $C$-index curves (mean (sd)), as obtained from the meta-regression models described in Section \ref{sec:2}. All areas were divided by the interval lengths $(\max_k (\tau_k)-\min_k (\tau_k))$ and multiplied by 1000.}
\begin{center}
\resizebox{\textwidth}{!}{
\begin{tabular}{l|rrr|rrr|rrr}
    \hline
    &\multicolumn{3}{c}{$\sigma_a=0$}&\multicolumn{3}{c}{$\sigma_a=0.01$}&\multicolumn{3}{c}{$\sigma_a=0.03$}\\
    \hline
     &short&moderate&long & short&moderate&long &short&moderate&long \\ 
\hline
MA (id) & 8.9 (4.9) & 10.0 (2.0) & 12.0 (1.5) & 9.1 (5.1) & 10.1 (2.1) & 13.4 (2.5) & 10.2 (6.1) & 10.9 (3.3) & 13.4 (2.5) \\ 
MA (id, last 50 \%) & 8.4 (4.0) & 11.7 (3.2) & 11.7 (1.6) & 8.6 (4.4) & 11.9 (3.6) & 14.1 (4.5) & 10.7 (6.4) & 13.4 (5.7) & 14.1 (4.5) \\ 
MA (id, last 30 \%) & 9.4 (5.0) & 13.5 (4.4) & 12.5 (2.5) & 9.8 (5.6) & 13.9 (5.1) & 15.6 (5.8) & 12.6 (8.3) & 15.8 (8.0) & 15.6 (5.8) \\ 
    \hline
MA (logit) & 7.9 (3.6) & 10.8 (2.5) & 11.4 (1.1) & 8.2 (3.9) & 10.8 (2.7) & 12.9 (2.2) & 9.7 (5.5) & 11.5 (3.9) & 12.9 (2.2) \\ 
MA (logit, last 50 \%) & 9.2 (4.6) & 12.7 (3.6) & 11.9 (1.8) & 9.5 (4.9) & 12.9 (4.0) & 14.1 (4.5) & 11.4 (6.9) & 14.1 (6.2) & 14.1 (4.5) \\ 
MA (logit, last 30 \%) & 10.4 (5.6) & 14.4 (4.8) & 12.7 (2.6) & 10.8 (6.1) & 14.7 (5.5) & 15.5 (5.8) & 13.5 (8.7) & 16.4 (8.3) & 15.5 (5.8) \\ 
    \hline
MA (asin) & 8.1 (4.2) & 10.2 (2.1) & 11.6 (1.3) & 8.3 (4.4) & 10.3 (2.2) & 13.1 (2.3) & 9.6 (5.6) & 10.9 (3.3) & 13.1 (2.3) \\ 
MA (asin, last 50 \%) & 8.6 (4.2) & 12.2 (3.4) & 11.8 (1.7) & 8.9 (4.5) & 12.4 (3.8) & 14.1 (4.5) & 10.9 (6.5) & 13.6 (5.9) & 14.1 (4.5) \\ 
MA (asin, last 30 \%) & 9.8 (5.3) & 13.9 (4.6) & 12.6 (2.5) & 10.2 (5.8) & 14.3 (5.3) & 15.5 (5.8) & 12.9 (8.4) & 16.0 (8.1) & 15.5 (5.8) \\ 
    \hline
linear (id) & 13.4 (7.6) & 10.0 (6.2) & 7.0 (2.2) & 13.3 (7.6) & 10.2 (6.2) & 9.7 (3.8) & 13.5 (7.9) & 11.5 (6.4) & 9.7 (3.8) \\ 
linear (logit) & 9.5 (5.7) & 7.4 (4.3) & 6.9 (1.6) & 9.7 (5.9) & 7.8 (4.4) & 9.5 (3.6) & 11.4 (6.8) & 10.2 (5.4) & 9.5 (3.6) \\ 
linear (asin) & 11.2 (6.7) & 8.4 (5.2) & 6.8 (1.7) & 11.3 (6.8) & 8.7 (5.3) & 9.6 (3.7) & 12.1 (7.2) & 10.6 (5.8) & 9.6 (3.7) \\ 
    \hline
RCS (id) & 16.0 (6.9) & 12.6 (5.7) & 6.9 (3.0) & 16.2 (6.8) & 12.9 (5.7) & 11.5 (4.4) & 17.6 (7.2) & 15.1 (6.0) & 11.5 (4.4) \\ 
RCS (logit) & 13.4 (5.8) & 10.8 (4.6) & 6.2 (2.5) & 13.7 (5.9) & 11.2 (4.6) & 11.2 (4.3) & 16.0 (6.6) & 14.1 (5.5) & 11.2 (4.3) \\ 
RCS (asin) & 14.5 (6.4) & 11.5 (5.1) & 6.4 (2.7) & 14.8 (6.4) & 11.9 (5.1) & 11.3 (4.3) & 16.6 (6.8) & 14.5 (5.7) & 11.3 (4.3) \\ 
    \hline
FP2 (id) & 14.4 (6.9) & 11.2 (5.6) & 6.3 (2.7) & 14.6 (6.9) & 11.5 (5.6) & 9.8 (4.1) & 15.5 (7.3) & 13.2 (6.0) & 9.8 (4.1) \\ 
FP2 (logit) & 11.4 (5.7) & 9.3 (4.4) & 5.7 (2.4) & 11.7 (5.9) & 9.7 (4.5) & 9.7 (4.1) & 13.6 (6.7) & 12.2 (5.3) & 9.7 (4.1) \\ 
FP2 (asin) & 12.7 (6.3) & 10.0 (5.0) & 5.9 (2.5) & 12.9 (6.4) & 10.4 (5.0) & 9.7 (4.1) & 14.3 (7.0) & 12.6 (5.6) & 9.7 (4.1) \\ 
\end{tabular}}
\end{center}
\label{tab:area}
\end{table}

In summary, our simulation study suggests that (i) spline and fractional polynomial meta-regression should be preferred over the exponential decay approach to model nonlinearities in $C(\tau)$, (ii) pooled estimates obtained from time-independent meta-analysis are reasonable approximations of $C(\tau )$ as long as follow-up times short, whereas it is necessary to consider increasingly complex meta-regression models with increasing $\tau$, and (iii) models with logit-transformed $C$-index estimates showed the overall best performance (compared to untransformed and arcsine-square-root-transformed estimates).

We further note that, compared to spline meta-regression, fractional polynomial meta-regression showed a slightly better overall performance in terms of the areas enclosed between the true and the fitted $C$-index curves. Importantly, the performance of fractional polynomial meta-regression could be improved further by optimizing the power values $p_1$ and $p_2$ (instead of considering the fixed values $-0.5$ and $0.5$, as done in this section). We will investigate this issue further in Section \ref{sec:4}. For details on variable and power selection in fractional polynomial regression, see Royston and Sauerbrei\cite{mfpbook}.

\section{Illustration}
\label{sec:4}

To illustrate the proposed methods, we analyzed data from the German Chronic Kidney Disease (GCKD) Study, which is an ongoing multi-center cohort study that enrolled 5,217 patients with chronic kidney disease (CKD). The aim of the study is to identify risk factors associated with CKD progression, cardiovascular events and death. For details on the inclusion/exclusion criteria and the design of the study, see Eckardt et al.\cite{eckardt}. Baseline data collection took place between March 2010 and March 2012; it comprised measurements on clinical and lifestyle variables (e.g.\@ coronary heart disease, smoking) and biomarker measurements obtained from blood and urine samples. Follow-up data are collected annually. The laboratory measurements collected for the GCKD Study have been used previously for predictive modeling and score development \citep{zacharias}.

An important characteristic of the GCKD Study is its wide geographical coverage. Altogether, there are nine study centers, each representing a specific German region with a distinct patient population. During the past years, it has become increasingly popular to account for such heterogeneity by synthesizing center-specific estimates via meta-analysis techniques\cite{franceschini, beekman, jogi, collatuzzo}. Here we followed this approach and used the GCKD data to evaluate a prognostic model in each of the nine centers, illustrating our proposed methodology by meta-analyzing the respective center-specific $C$-index estimates ($K=9$). Note that the availability of individual patient data allowed us to estimate $C(\tau )$ in each center at arbitrary time horizons.

\begin{figure}
\centering
\includegraphics[width=14cm]{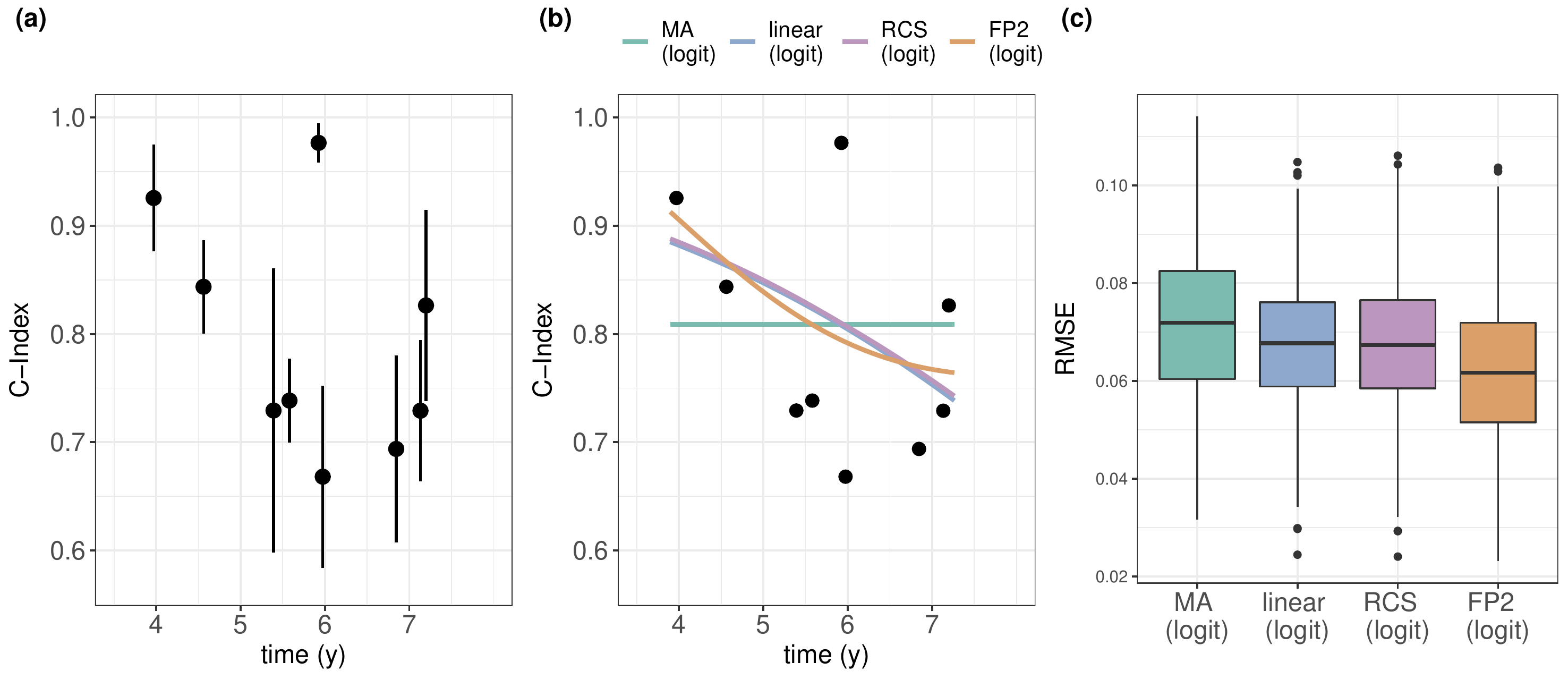}
\caption[Figure2f]{Analysis of the GCKD Study data. The left plot shows the nine center-specific $C$-index estimates at randomly generated truncation times $\tau_k$, $k=1,\ldots , 9$. Estimates were obtained by application of Uno's $C$ to the GCKD analysis data. The colored lines in the middle plot refer to the back-transformed meta-analysis and -regression curves obtained by fitting the models of Section~\ref{sec:24} to the logit-transformed $C$-index estimates (MA = standard random effects meta-analysis ignoring time dependency, linear = linear meta-regression, RCS = restricted cubic spline meta-regression, FP2 = fractional polynomial meta-regression). The right plot shows the RMSE values obtained from the bootstrapped test data (1000 replications).}
\label{fig:gckd}
\end{figure}

For model building and evaluation we considered the endpoint ``time to cardiovascular death''. Data were exported from the GCKD database after the 8th follow-up examination (maximum follow-up time 2,933 days, median = 2,554 days, first quartile = 2,060 days, third quartile = 2,591 days, cardiovascular death rate = 200/4,455 = $4.5\%$ after listwise deletion of patients with a missing value in at least one of the covariates). In the first step, we split the data randomly into three equally sized parts: The first part was used as {\it training data} for model building, the second part was used as {\it analysis data} for prediction and meta-regression, and the third part was used as {\it test data} for evaluating the performance of the meta-regression models. In the second step, we derived a prediction model for cardiovascular death by fitting a Cox regression model to the training data. The following (pre-selected) baseline covariates were included in the model:  C-reactive protein (mg/L), cholesterol (mg/dL), calcium (mmol/L), phosphate (mmol/L), albumin (g/L), cystatin C (mg/L), age (years), sex (male/female), urine albumin-to-creatinine ratio (mg/g), hypertension (yes/no), previous coronary heart disease (yes/no), smoking (non-smoker, former smoker, current smoker), and estimated glomerular filtration rate (mL/min/1.73 m$^2$). Furthermore, we generated a random truncation time $\tau_k$, $k=1,\ldots , 9$, for each of the study centers, restricting $\tau_k$ to be larger than 3.5 years in order to obtain sufficiently large event counts. In the third step, we used the coefficients of the Cox model to predict the values of $\eta$ in the analysis data. These values were subsequently used to estimate the restricted concordance probability in each study center at the center-specific truncation times $\tau_k$. The resulting estimates $\hat{C}_k$ (obtained by application of Uno's $C$) are visualized in Figure~\ref{fig:gckd}(a). In the fourth step, we meta-analyzed the center-specific $C$-index estimates by applying the methods presented in Section~\ref{sec:24} to the pairs of values $(\tau_1, \hat{C}_1), \ldots , (\tau_9, \hat{C}_9)$. Based on the results of our simulation study, $C$-index estimates were logit-transformed before model fitting. In the fifth step, we generated 1000 bootstrap samples from the test data and re-estimated the concordance probabilities $C(\tau_k )$, $k=1,\ldots , 9$, as well as the fitted values $g^{-1}(f(\tau_k ; \hat{\gamma}))$ (obtained from meta-regression) in each of the samples. Furthermore, we computed the weighted root mean squared error (RMSE, defined by $\big[ \sum_{k=1}^K n_k/n \, ( \hat{C}_k - g^{-1}(f(\tau_k ; \hat{\gamma})) )^2 \big]^{1/2}$), which was used to evaluate and compare the performance of the meta-regression models.

The fitted curves obtained from the analysis data are presented in Figure~\ref{fig:gckd}(b). It is seen that the meta-regression models (accounting for the length of follow-up) resulted in very similar fits. Fractional polynomial meta-regression seemed to perform best by visual inspection. The model summaries (given in Table \ref{tab:gckdresults}) confirm these results, with the values of the estimated between-study standard deviation $\hat{\sigma}_a$ ranging between 0.708 and 0.821 on the logit scale. 
Boxplots of the RMSE values (obtained from the bootstrapped test data) are shown in Figure \ref{fig:gckd}(c). Again, it can be seen that the models performed similarly, with the highest RMSE value observed for the meta-analysis model and the lowest RMSE value for the fractional polynomial meta-regression model.

\begin{table}[ht]
\caption{Analysis of the GCKD Study data. The table summarizes the fits of the meta-analysis and -regression models, as obtained by applying the methods of Section \ref{sec:24} to the center-specific $C$-index values (estimated from the GCKD analysis data). The logistic transformation was applied to the estimated $C$-index values before model fitting. The table presents the values of the estimated between-study standard deviation $\hat{\sigma}_a$ (on the logit scale), the test statistic for residual heterogeneity $Q$ (following a Chi-squared distribution under the null hypothesis of homogeneous residuals), its degrees of freedom (df), and the corresponding p-value.}
\begin{center}
\begin{tabular}{rlrrrrrr}
  \hline
  & $\hat{\sigma}_a$ & $Q$ & df & p-value \\ 
  \hline
    MA (logit)      & 0.708 & 22.3 & 8 & 0.0043 \\ 
    linear (logit)  & 0.711 & 19.9 & 7& 0.0059 \\ 
    RCS (logit)     & 0.821 & 18.9 & 6& 0.0043 \\ 
    FP2 (logit)    &  0.793 & 18.1 & 6& 0.0060 \\ 
   \hline
\end{tabular}
\end{center}
\label{tab:gckdresults}
\end{table}

\begin{figure}
\centering
\includegraphics[width=14cm]{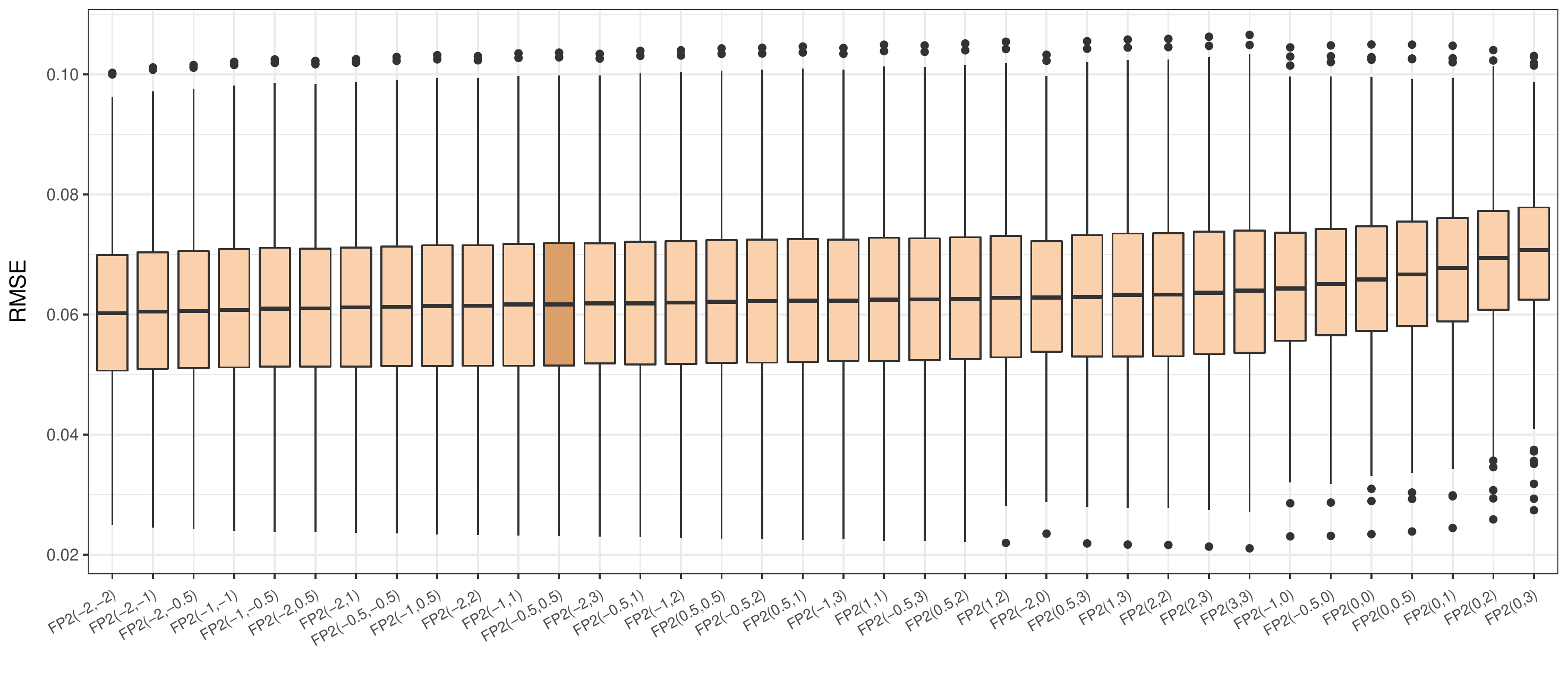}
\caption[Figure2f]{Analysis of the GCKD data. The boxplots show the RMSE values obtained from the bootstrapped test data (1000 replications) when evaluating all possible combinations of the power values $p_1, p_2 \in \{-2, -1, -0.5, 0, 0.5, 1, 2, 3\}$ of the FP2 models. The boxplots are ordered by median RMSE value. The dark orange boxplot corresponds to the powers of the FP2 model from Section \ref{sec:24} ($p_1=-0.5$, $p_2=0.5$).}
\label{fig:gckdFP}
\end{figure}

In the final step, we investigated whether we could improve the performance of fractional polynomial meta-regression by optimizing the power values of the FP2 model. To this purpose, we repeated the bootstrap analysis of the GCKD test data, this time computing the RMSE values obtained from all possible combinations of the powers $p_1, p_2 \in \{-2, -1, -0.5, 0, 0.5, 1, 2, 3\}$. The results of our analysis suggest that the RMSE values were not very sensitive to the choice of powers (Figure \ref{fig:gckdFP}). In particular, the performance of our initial model from Section \ref{sec:24} ($p_1=-0.5$, $p_2=0.5$, Figure \ref{fig:gckd}(c)) was close to the performance of the optimal model with $p_1=p_2=-2$.

\section{Discussion}

The development of prognostic models has become a predominant task in medical and epidemiological research. As noted by Riley et al.\cite{rileyPrognostic}, ``prognostic factors have many potential uses, including aiding treatment and lifestyle decisions, improving individual risk prediction, providing novel targets for new treatment, and enhancing the design and analysis of randomised trials identify patients for trials``. To this end, a large number of prognostic scores has been developed, requiring proper validation to become accepted for eventual use in clinical practice. The gold standard for validation is to analyze the performance of novel scores using large external cohorts; however, for a variety of reasons (including confidentiality and logistical issues), this is often not possible. It is therefore important to combine the results of smaller validation studies using meta-analysis techniques. 

In this paper we proposed and evaluated a framework for meta-analyzing the concordance index for time-to-event data, which has become an established measure for the discriminative ability of prognostic scores (see Steyerberg\cite{Steyerberg2019} for a comprehensive introduction to predictive modeling, also including other aspects of validation like calibration and clinical usefulness). We analyzed the inherent time-dependency of $C$-index estimates (noted previously by Longato et al.\cite{longato}) and proposed methods to account for this time-dependency using meta-regression models. In this respect, our paper connects to Debray et al.\cite{debrayGuide} who noted that ``[...]  researchers often refrain from undertaking a quantitative synthesis or meta-analysis of the predictive performance of a specific model. Potential reasons for this pitfall are [...] or simply a lack of methodological guidance.''

A key result of this work is that meta-regression models including the study-specific truncation time as covariate perform systematically better than classical random-effects meta-analysis when follow-up times of validation studies are long. Conversely, pooled estimates obtained from classical meta-analysis are reasonable approximations of the restricted concordance probability as long as follow-up times short. We acknowledge that, in practice, it might be challenging to determine whether a time horizon should be considered ``short'' or ``long'', in particular when observation times are affected by high drop-out rates and/or the presence of competing events. Still, we recommend to carefully investigate this issue, especially since $C$-index values often tend to decrease with $\tau$, implying that studies with a short follow-up time might suggest an overly optimistic discrimination accuracy. Generally, the rate of administrative censoring might be an indicator of whether follow-up times might be considered ``long'' or ``short''. Furthermore, we recommend visual inspection of $C$-index estimates in order to examine their dependency on $\tau$.

Based on our numerical experiments, we recommend to transform $C$-index values using a logistic transformation and to employ either restricted cubic splines or fractional polynomials to model the functional relationship between the truncation time and the concordance index. We further recommend to prefer fractional polynomials over splines in settings where the number of studies is ``small'' ($5 \le K \le 10$), as they typically involve fewer degrees of freedom than restricted cubic splines. In case of convergence problems (which might become an issue when the number of studies is smaller than five), our framework readily allows for switching to a simpler model (e.g.\@ a linear meta-regression model). We also note that our proposed models could be extended by additional covariates reflecting different inclusion criteria in the analyzed studies. Along the same lines, our framework could be adapted to models with competing events\cite{vanGelovenBMJ}.

A key barrier to meta-analyzing $C$-index values is the huge variety of estimators that have been proposed during the past decades (such as Harrell's $C$ and Uno's $C$)\cite{schmidPotapov}. Since each of these estimators comes with a different set of assumptions and/or properties, it is challenging to synthesize validation studies relying on different kinds of estimators. Importantly, some of the estimators are known to be systematically biased, e.g.\@ when they rely on a Cox model (whose assumptions might be violated) or when they show a censoring bias (such as Harrell's $C$). We argue that these systematic deviations should not be represented in a meta-regression model by zero-mean random effects. Instead, we suggest to develop methodological guidance on the definition and use of appropriate estimators for the evaluation of discriminatory power, aiming at a unified methodology that would become a standard in future validation studies. Work on such guidance is e.g.\@ undertaken by the STRengthening Analytical Thinking for Observational Studies (STRATOS) initiative \citep{sauerbrei2014stratos}.

Meta-regression of $C$-index estimates is also compromised by the lack of proper reporting. In fact, when searching for a real-world application to be presented in Section \ref{sec:4}, we found that most published studies reporting $C$-index estimates did {\it not} include any information on the respective time horizon. In some cases, we were able to approximate this time horizon by the length of the respective follow-up time; however, in many cases the time horizon was not mentioned at all. Based on the findings presented in Section~\ref{sec:3} of this paper, we suggest to always report the time horizons together with $C$-index estimates in future validation studies. We further suggest to report and visualize the whole estimated $C$-index curve whenever a meta-regression has been performed.

We finally note that the concordance index is (by far) not the only prognostic measure to be affected by an inherent time dependency. Another important example are incidence rates, which by definition depend on the time frames under consideration. Clearly, the lengths of these time frames have to be considered when meta-analyzing incidence rates (see Olaciregui-Dague et al.\cite{karmele} for a recent example). Further research is needed to evaluate possible adaptions of our methodology to these measures.

\subsection*{Acknowledgements}

We thank the GCKD Study Investigators for providing data of the GCKD Study for illustrative purposes. The GCKD study was supported by a grant from the KfH Foundation for Preventive Medicine (https://www.kfh-stiftung-praeventivmedizin.de). Tim Friede is grateful for support by the Volkswagen Foundation (Az.: 98 948; ``Bayesian and Nonparametric Statistics-Teaming up two opposing theories for the benefit of prognostic studies in Covid-19'').

\bibliography{wileyNJD-AMA}%

\newpage
\raggedbottom
\newcommand{\beginsupplement}{%
        \setcounter{table}{0}
        \renewcommand{\thetable}{S\arabic{table}}%
        \setcounter{figure}{0}
        \renewcommand{\thefigure}{S\arabic{figure}}%
     }

\authormark{Supporting Information - Matthias Schmid \textsc{et al}}

%\maketitle
\beginsupplement

\vspace*{.5cm}
\Large \noindent
{Supporting Information}

\normalsize

\section*{} 

\begin{figure}[h!]
\centering
\includegraphics[width=10cm]{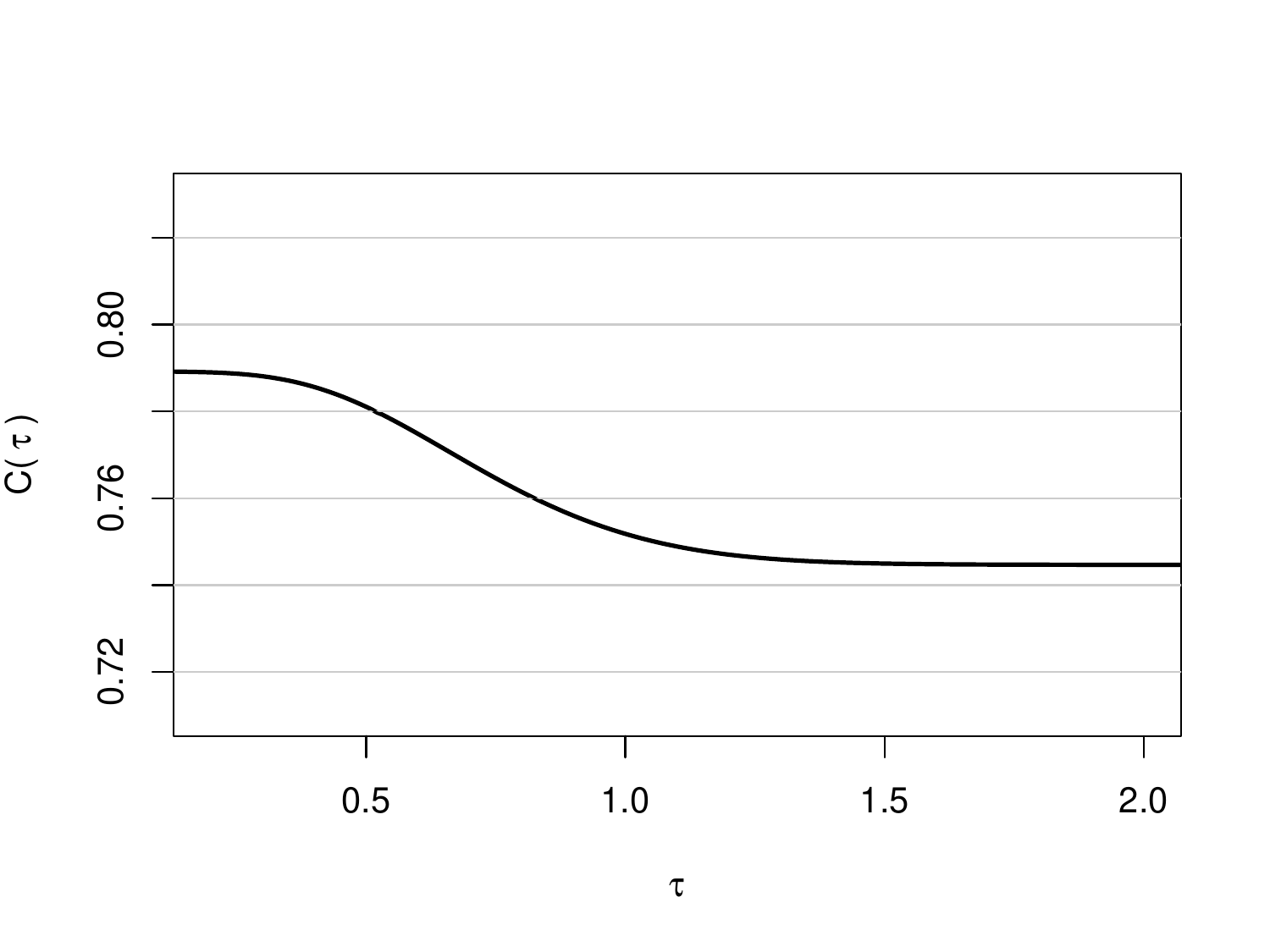}
\caption[A1]{$C$-index curve of the Weibull model used for the simulation study in Section 3 of the paper.}
\label{figApp1:trueCindex}
\end{figure}

\newpage

\begin{figure}[h!]
\centering
\includegraphics[width=12cm]{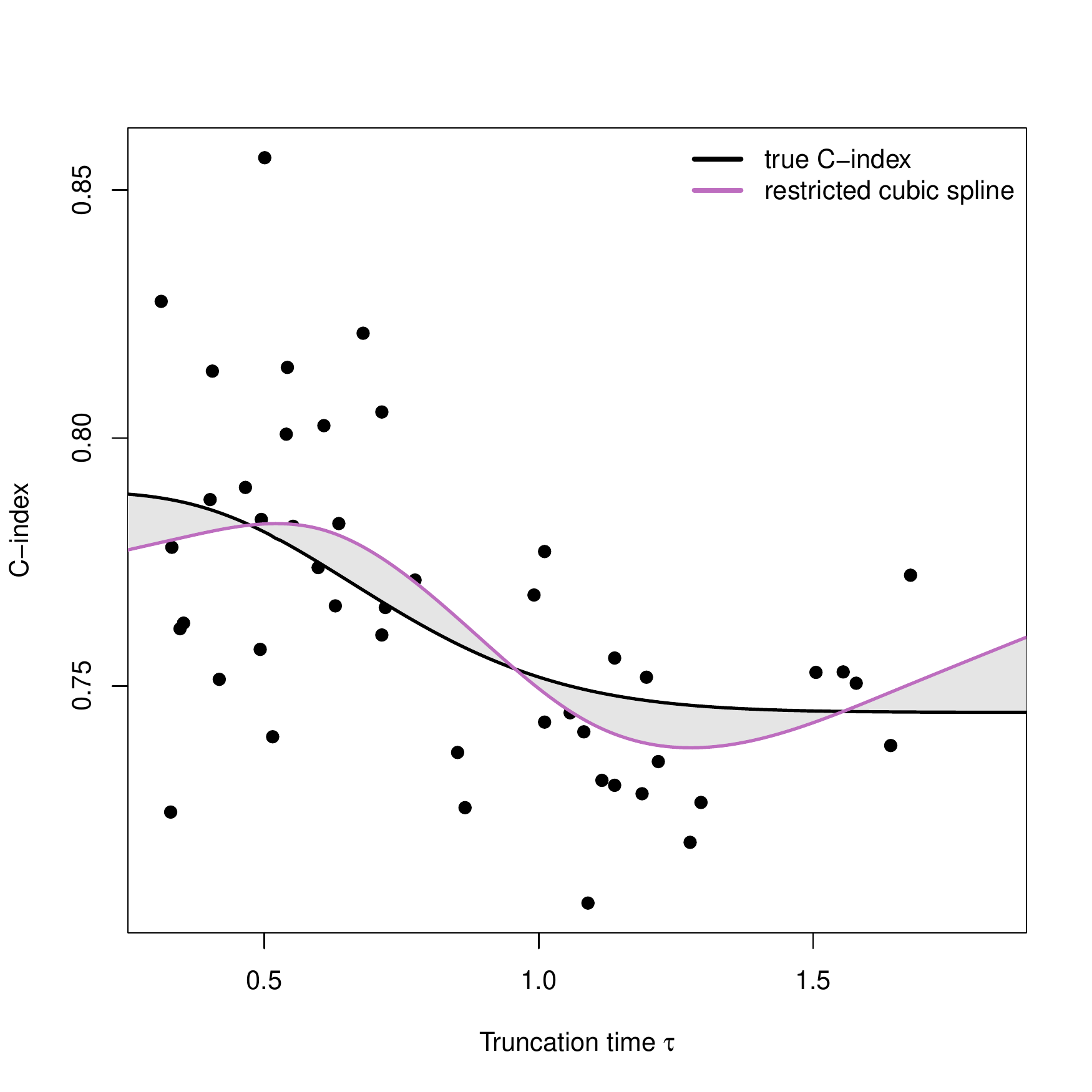}
\caption[Figure2b]{Example of an area enclosed by the true and the estimated $C$-index curves (black line = true $C$-index, pink line = restricted cubic spline estimate). The size of the gray area (divided by the length of the time interval) was used as an evaluation measure for the simulation study in Section 3 of the paper.}
\label{fig2b:aucexample_logit}
\end{figure}

\newpage

% failure rates K=15
\begin{table}[!ht]
\caption{Results of the simulation study ($K=15$). The table summarizes the failure rates ($\%$) of the meta-regression models described in Section 2 of the paper. Failure rates were defined by the percentages of simulation runs in which the respective R~fitting functions issued either an error or a warning (MA = meta-analysis, linear = linear meta-regression, RCS = restricted cubic spline meta-regression, FP2 = 2nd degree fractional polynomial meta-regression).}
\begin{center}
\resizebox{0.8\textwidth}{!}{
\begin{tabular}{l|rrr|rrr|rrr}
    \hline
    &\multicolumn{3}{c}{$\sigma_a=0$}&\multicolumn{3}{c}{$\sigma_a=0.01$}&\multicolumn{3}{c}{$\sigma_a=0.03$}\\   
    \hline
     &short&moderate&long & short&moderate&long &short&moderate&long \\ 
\hline MA (id) & 0.1 & 0.1 & 0.2 & 0.3 & 0.3 & 0.7 & 0.2 & 0.1 & 0.0 \\ 
  MA (id, last 50 \%) & 0.0 & 0.3 & 0.3 & 0.1 & 0.4 & 0.1 & 0.1 & 0.2 & 0.1 \\ 
  MA (id, last 30 \%) & 0.1 & 0.2 & 0.4 & 0.0 & 0.2 & 0.2 & 0.1 & 0.0 & 0.1 \\ \hline
  MA (logit) & 0.4 & 0.5 & 0.1 & 0.4 & 0.4 & 0.0 & 0.2 & 0.2 & 0.0 \\ 
  MA (logit, last 50 \%) & 0.1 & 0.2 & 0.0 & 0.3 & 0.3 & 0.0 & 0.1 & 0.1 & 0.0 \\ 
  MA (logit, last 30 \%) & 0.1 & 0.0 & 0.0 & 0.1 & 0.0 & 0.0 & 0.0 & 0.0 & 0.0 \\ \hline
  MA (asin) & 0.2 & 0.3 & 0.2 & 0.1 & 0.3 & 0.1 & 0.0 & 0.1 & 0.0 \\ 
  MA (asin, last 50 \%) & 0.0 & 0.1 & 0.2 & 0.0 & 0.3 & 0.1 & 0.1 & 0.0 & 0.1 \\ 
  MA (asin, last 30 \%) & 0.0 & 0.2 & 0.2 & 0.0 & 0.0 & 0.2 & 0.2 & 0.1 & 0.0 \\ \hline
  linear (id) & 0.3 & 0.5 & 0.3 & 0.5 & 0.4 & 0.6 & 0.3 & 0.0 & 0.0 \\ 
  linear (logit) & 0.7 & 0.8 & 0.2 & 0.5 & 0.6 & 0.1 & 0.4 & 0.4 & 0.0 \\ 
  linear (asin) & 0.2 & 0.5 & 0.2 & 0.3 & 0.4 & 0.2 & 0.3 & 0.3 & 0.0 \\ \hline
  RCS (id) & 0.3 & 0.0 & 0.0 & 0.3 & 0.4 & 0.1 & 0.0 & 0.3 & 0.0 \\ 
  RCS (logit) & 0.6 & 0.8 & 0.0 & 0.5 & 0.6 & 0.2 & 0.1 & 0.3 & 0.0 \\ 
  RCS (asin) & 0.3 & 0.2 & 0.1 & 0.2 & 0.5 & 0.5 & 0.0 & 0.2 & 0.0 \\ \hline
  FP2 (id) & 0.4 & 0.4 & 0.2 & 0.2 & 0.1 & 0.0 & 0.0 & 0.2 & 0.0 \\ 
  FP2 (logit) & 0.6 & 0.7 & 0.0 & 0.4 & 0.7 & 0.3 & 0.1 & 0.1 & 0.0 \\ 
  FP2 (asin) & 0.2 & 0.3 & 0.3 & 0.2 & 0.2 & 0.3 & 0.0 & 0.4 & 0.0 \\ \hline
  exponential decay (id) & 59.2 & 52.5 & 27.4 & 61.6 & 48.6 & 28.1 & 62.8 & 50.2 & 37.8 \\ 
  exponential decay (logit) & 71.0 & 62.3 & 33.6 & 74.4 & 63.5 & 34.7 & 76.4 & 66.6 & 47.8 \\ 
  exponential decay (asin) & 65.7 & 54.0 & 29.2 & 65.7 & 51.9 & 30.5 & 68.2 & 55.6 & 43.7 \\ 
\hline
\end{tabular}}
\end{center}
\label{tab:simresults_errwarn2}
\end{table}

\newpage

% failiure rates, K=50
\begin{table}[!ht]
\caption{Results of the simulation study ($K=50$). The table summarizes the failure rates ($\%$) of the meta-regression models described in Section 2 of the paper. Failure rates were defined by the percentages of simulation runs in which the respective R~fitting functions issued either an error or a warning.}
\begin{center}
\resizebox{0.8\textwidth}{!}{
\begin{tabular}{l|rrr|rrr|rrr}
    \hline
    &\multicolumn{3}{c}{$\sigma_a=0$}&\multicolumn{3}{c}{$\sigma_a=0.01$}&\multicolumn{3}{c}{$\sigma_a=0.03$}\\
        \hline
     &short&moderate&long & short&moderate&long &short&moderate&long \\ 
\hline MA (id) & 0.0 & 0.0 & 0.1 & 0.0 & 0.0 & 0.0 & 0.0 & 0.0 & 0.0 \\ 
  MA (id, last 50 \%) & 0.0 & 0.0 & 0.1 & 0.0 & 0.0 & 0.0 & 0.0 & 0.1 & 0.0 \\ 
  MA (id, last 30 \%) & 0.0 & 0.1 & 0.1 & 0.0 & 0.0 & 0.0 & 0.0 & 0.0 & 0.0 \\ \hline 
  MA (logit) & 0.1 & 0.0 & 0.0 & 0.1 & 0.0 & 0.0 & 0.0 & 0.0 & 0.0 \\ 
  MA (logit, last 50 \%) & 0.2 & 0.0 & 0.0 & 0.2 & 0.0 & 0.0 & 0.1 & 0.0 & 0.0 \\ 
  MA (logit, last 30 \%) & 0.1 & 0.0 & 0.0 & 0.0 & 0.1 & 0.0 & 0.1 & 0.0 & 0.0 \\ \hline 
  MA (asin) & 0.1 & 0.2 & 0.0 & 0.0 & 0.0 & 0.1 & 0.0 & 0.0 & 0.0 \\ 
  MA (asin, last 50 \%) & 0.0 & 0.0 & 0.0 & 0.1 & 0.1 & 0.0 & 0.0 & 0.0 & 0.0 \\ 
  MA (asin, last 30 \%) & 0.0 & 0.0 & 0.1 & 0.0 & 0.1 & 0.0 & 0.0 & 0.0 & 0.0 \\ \hline 
  linear (id) & 0.1 & 0.3 & 0.1 & 0.1 & 0.0 & 0.1 & 0.0 & 0.0 & 0.0 \\ 
  linear (logit) & 0.1 & 0.0 & 0.0 & 0.1 & 0.0 & 0.0 & 0.3 & 0.1 & 0.1 \\ 
  linear (asin) & 0.1 & 0.1 & 0.0 & 0.0 & 0.1 & 0.0 & 0.3 & 0.1 & 0.1 \\ \hline 
  RCS (id) & 0.2 & 0.0 & 0.1 & 0.0 & 0.1 & 0.0 & 0.0 & 0.0 & 0.0 \\ 
  RCS (logit) & 0.2 & 0.0 & 0.0 & 0.0 & 0.0 & 0.0 & 0.3 & 0.1 & 0.1 \\ 
  RCS (asin) & 0.1 & 0.2 & 0.0 & 0.0 & 0.0 & 0.0 & 0.3 & 0.1 & 0.1 \\ \hline 
  FP2 (id) & 0.1 & 0.2 & 0.0 & 0.0 & 0.1 & 0.0 & 0.0 & 0.0 & 0.0 \\ 
  FP2 (logit) & 0.0 & 0.0 & 0.0 & 0.0 & 0.0 & 0.0 & 0.3 & 0.1 & 0.1 \\ 
  FP2 (asin) & 0.1 & 0.1 & 0.0 & 0.1 & 0.0 & 0.1 & 0.3 & 0.1 & 0.1 \\ \hline 
  exponential decay (id) & 51.7 & 38.6 & 10.3 & 51.2 & 33.6 & 10.5 & 57.3 & 42.8 & 15.8 \\ 
  exponential decay (logit) & 71.1 & 60.1 & 22.1 & 69.1 & 57.1 & 15.9 & 76.6 & 67.7 & 34.7 \\ 
  exponential decay (asin) & 60.0 & 45.5 & 12.0 & 59.5 & 41.8 & 12.6 & 65.6 & 53.5 & 29.2 \\ 
\hline
\end{tabular}}
\end{center}
\label{tab:simresults_errwarn3}
\end{table}

\newpage
%FIGURES 0.8 Cindex K=15

\begin{figure}[h!]
\centering
\includegraphics[width=14cm]{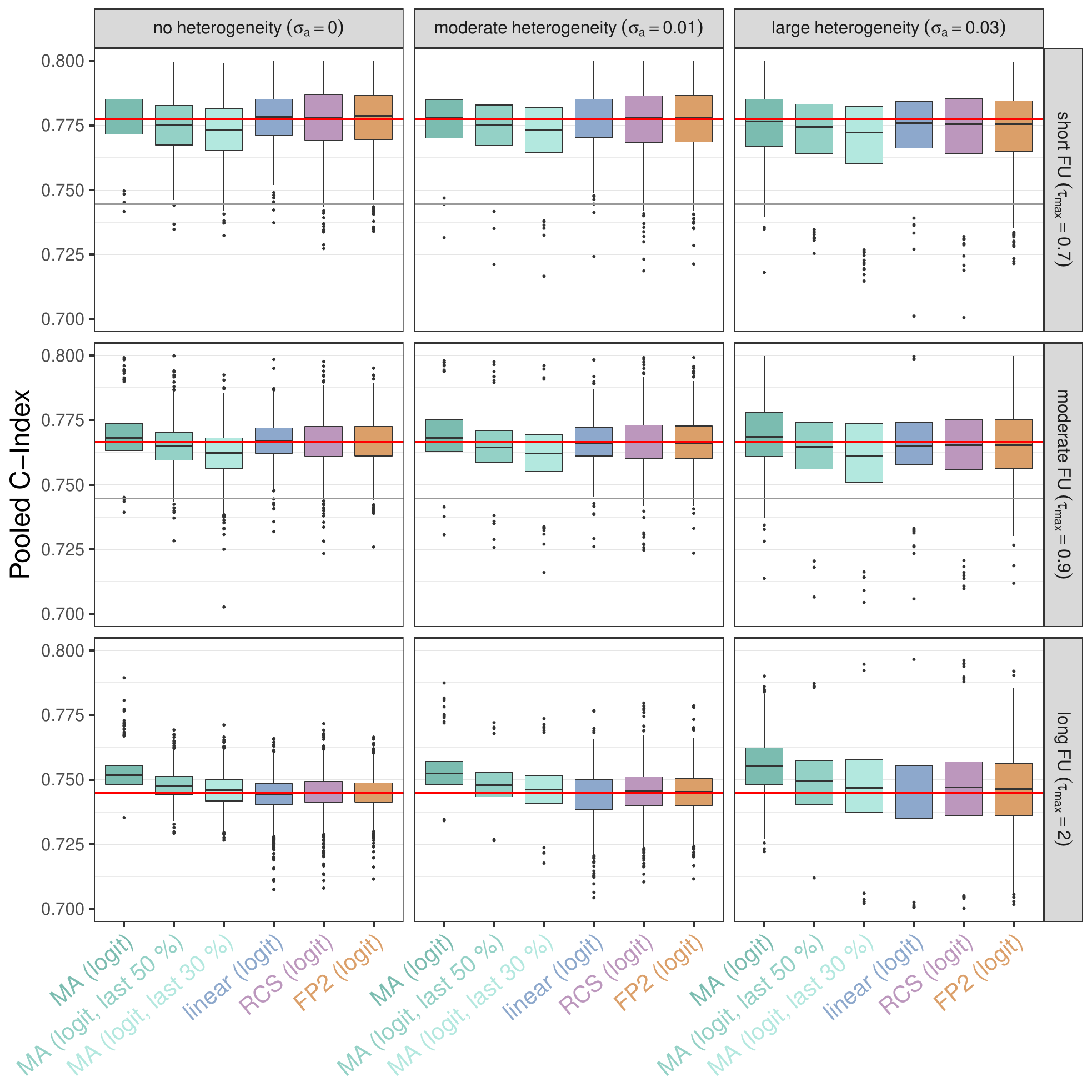}
\caption[Figure2b]{Results of the simulation study  ($K=15$). The boxplots summarize the pooled estimates of the restricted concordance index at $0.8\cdot\tau_{\text{max}}$. All $C$-index estimates were transformed using a logistic transformation before model fitting. The red and the black lines refer to the true values of $C(0.8\cdot\tau_{\text{max}})$ and the unrestricted values of the concordance index, respectively. Note that the black lines coincide with the red lines in the lower three panels.}
\label{fig2b:simresults_logit_K15}
\end{figure}
\newpage

\begin{figure}[h!]
\centering
\includegraphics[width=14cm]{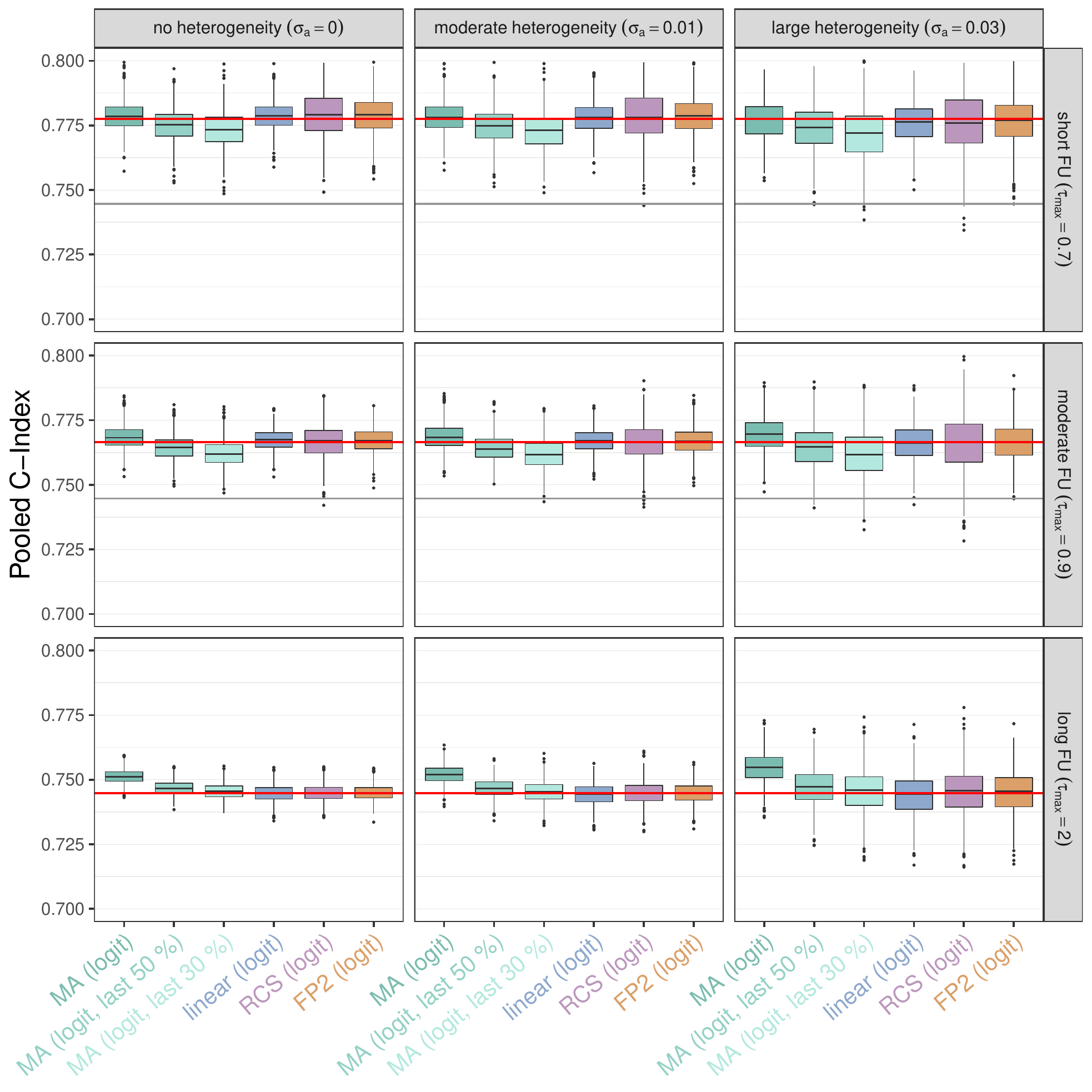}
\caption[Figure2b]{Results of the simulation study  ($K=50$). The boxplots summarize the pooled estimates of the restricted concordance index at $0.8\cdot\tau_{\text{max}}$. All $C$-index estimates were transformed using a logistic transformation before model fitting. The red and the black lines refer to the true values of $C(0.8\cdot\tau_{\text{max}})$ and the unrestricted values of the concordance index, respectively. Note that the black lines coincide with the red lines in the lower three panels.}
\label{fig2b:simresults_logit_K50}
\end{figure}

\newpage

\begin{figure}[h!]
\centering
\includegraphics[width=14cm]{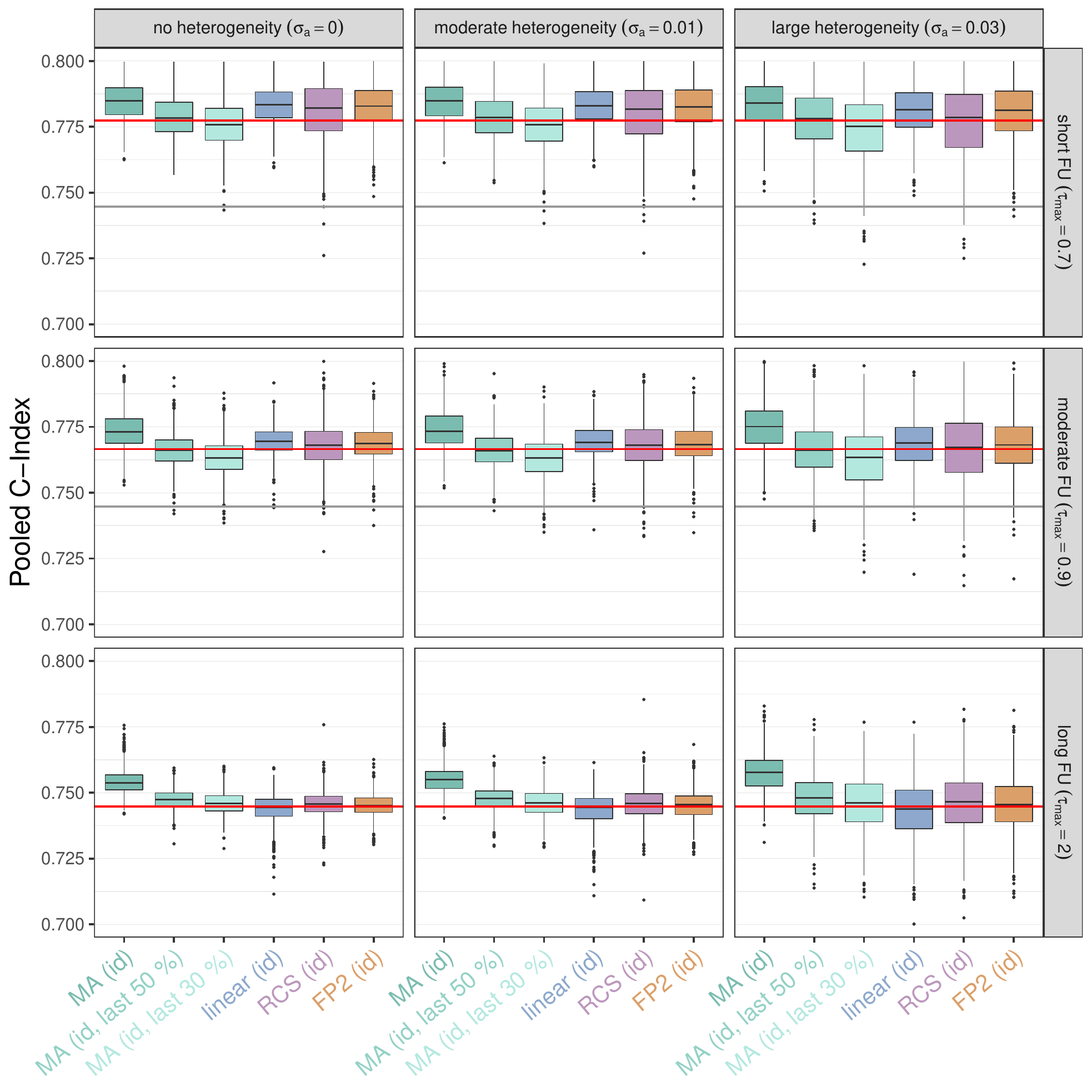}
\caption[Figure2b]{Results of the simulation study  ($K=30$). The boxplots summarize the pooled estimates of the restricted concordance index at $0.8\cdot\tau_{\text{max}}$. All $C$-index estimates were left untransformed during model fitting. The red lines refer to the true values of $C(0.8\cdot\tau_{\text{max}})$, and the black lines refer to the unrestricted values of the concordance index. Note that the black lines coincide with the red lines in the lower three panels.}
\label{fig2b:simresults_id}
\end{figure}

\newpage

\begin{figure}[h!]
\centering
\includegraphics[width=14cm]{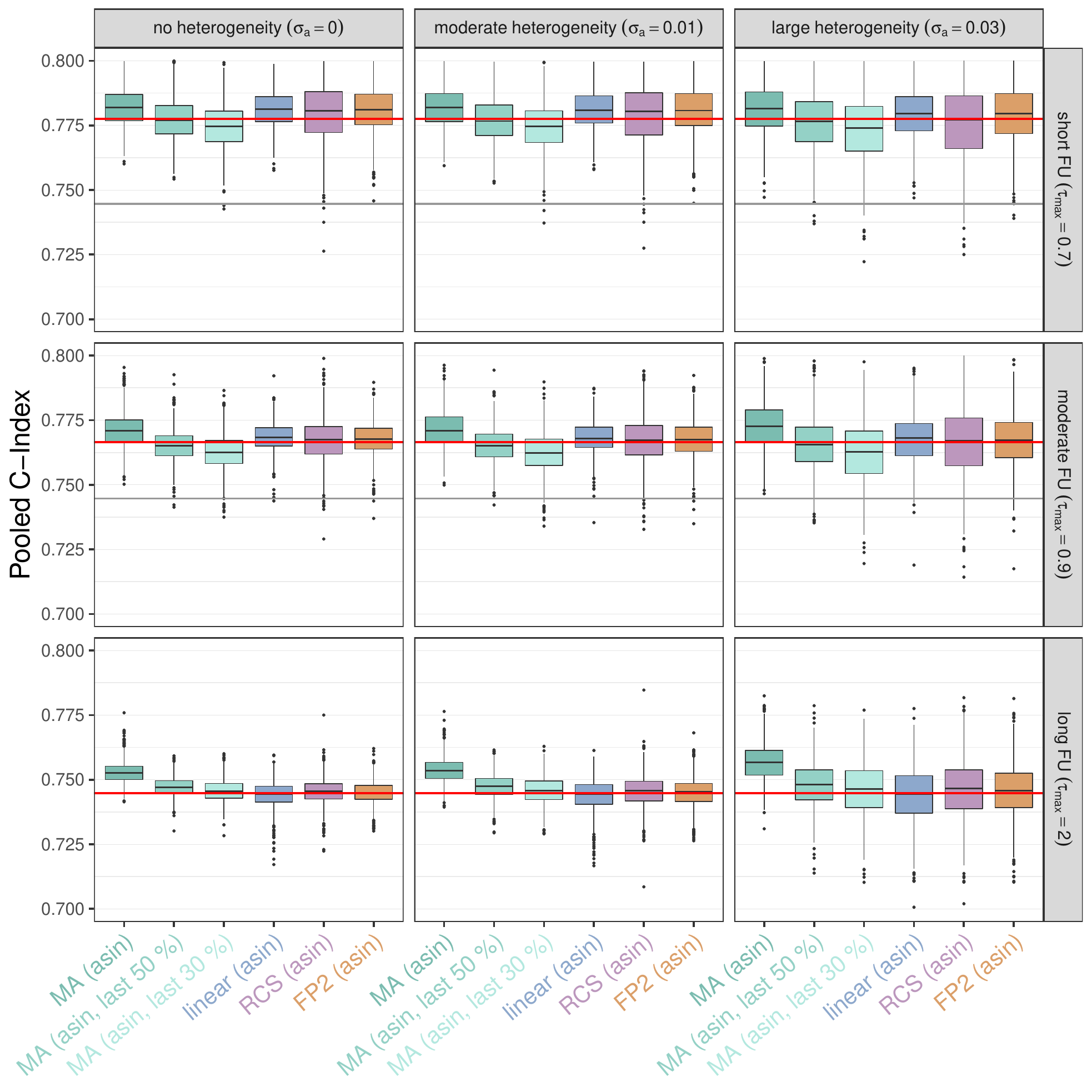}
\caption[Figure2b]{Results of the simulation study  ($K=30$). The boxplots summarize the pooled estimates of the restricted concordance index at $0.8\cdot\tau_{\text{max}}$. All $C$-index estimates were transformed using an arcsine-square-root transformation before model fitting. The red lines refer to the true values of $C(0.8\cdot\tau_{\text{max}})$, and the black lines refer to the unrestricted values of the concordance index. Note that the black lines coincide with the red lines in the lower three panels.}
\label{fig2b:simresults_asin}
\end{figure}

\newpage

\begin{figure}[h!]
\centering
\includegraphics[width=14cm]{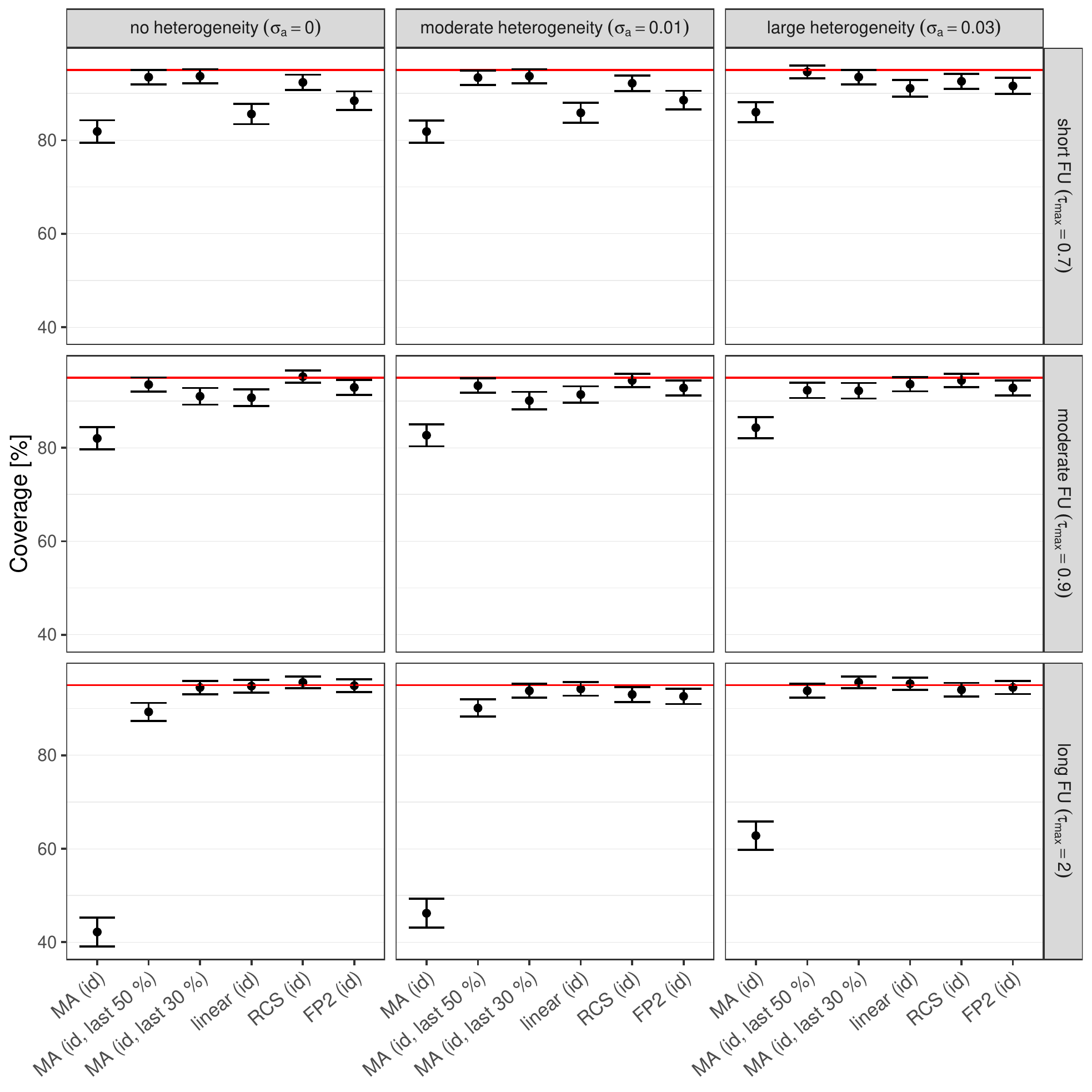}
\caption[Figure2b]{Results of the simulation study  ($K=30$). The boxplots show the estimated coverage probabilities (\%), i.e.\@ the proportion of simulation runs in which the 95\% Hartung-Knapp confidence intervals contained the true value of $C(0.8\cdot\tau_{\text{max}})$. Confidence limits (represented by the black lines) were computed as $[\hat{p} \pm 1.96 \cdot \sqrt{\hat{p} \cdot (1-\hat{p}) / 1000}]$, where $\hat{p}$ denotes the point estimate of the coverage probability. The red lines refer to the $95\%$ confidence level. Model fitting was based on the untransformed $C$-index estimates.}
\label{fig2b:coverage_id}
\end{figure}

\newpage

\begin{figure}[h!]
\centering
\includegraphics[width=14cm]{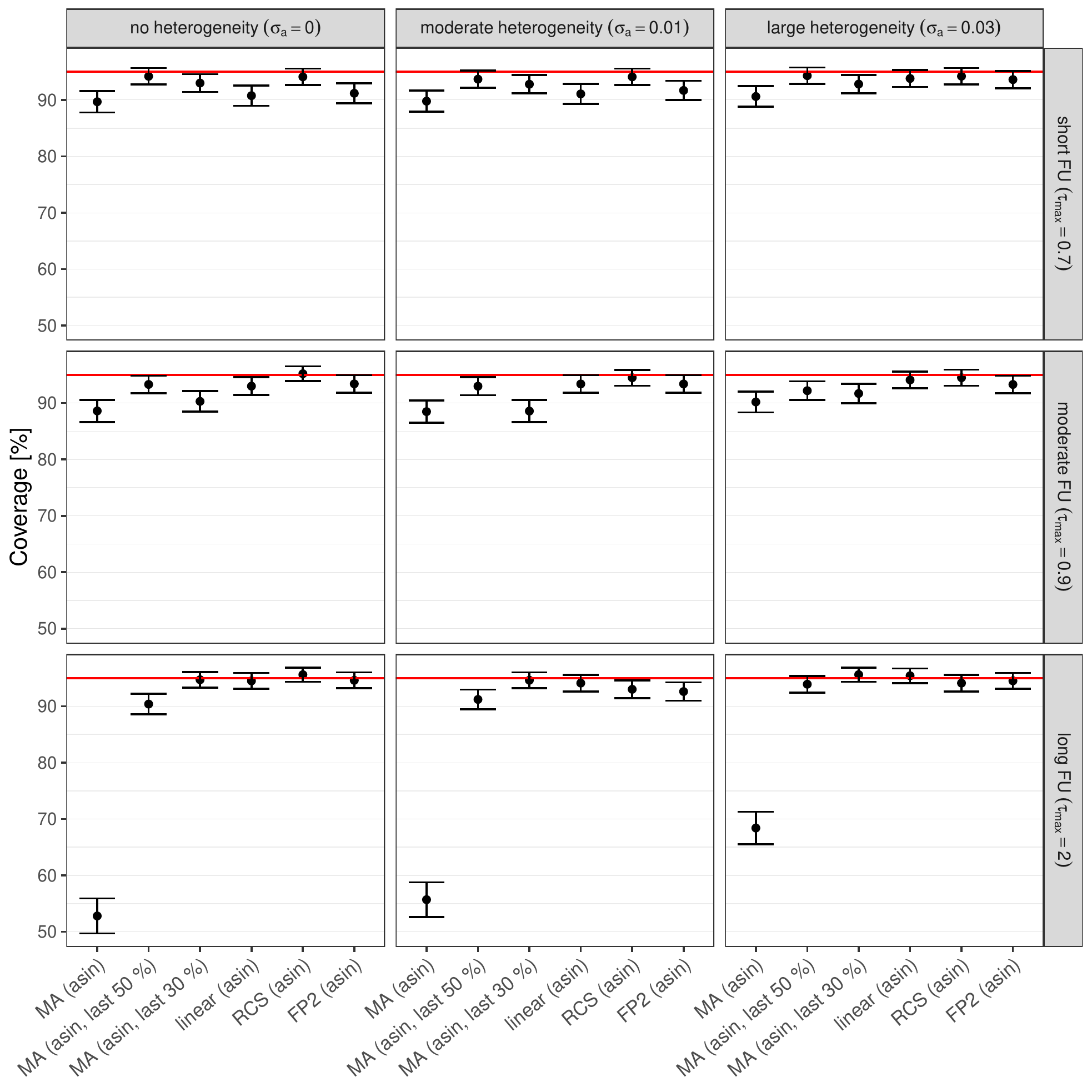}
\caption[Figure2b]{Results of the simulation study  ($K=30$). The boxplots show the estimated coverage probabilities (\%), i.e.\@ the proportion of simulation runs in which the 95\% Hartung-Knapp confidence intervals contained the true value of $C(0.8\cdot\tau_{\text{max}})$. Confidence limits (represented by the black lines) were computed as $[\hat{p} \pm 1.96 \cdot \sqrt{\hat{p} \cdot (1-\hat{p}) / 1000}]$, where $\hat{p}$ denotes the point estimate of the coverage probability. The red lines refer to the $95\%$ confidence level. All $C$-index estimates were transformed by an arcsine square root transformation before model fitting.}
\label{fig2b:coverage_asin}
\end{figure}

\newpage

\begin{table}[!ht]
\caption{Results of the simulation study ($K=15$). The table summarizes the areas enclosed by the true and the estimated $C$-index curves, as obtained from the meta-regression models described in Section 2 of the paper. All areas were divided by the interval lengths $(\max_k (\tau_k)-\min_k (\tau_k))$ and multiplied by 1000. }
\begin{center}
\resizebox{\textwidth}{!}{
\begin{tabular}{l|rrr|rrr|rrr}
    \hline
    &\multicolumn{3}{c}{$\sigma_a=0$}&\multicolumn{3}{c}{$\sigma_a=0.01$}&\multicolumn{3}{c}{$\sigma_a=0.03$}\\
        \hline
     &short&moderate&long & short&moderate&long &short&moderate&long \\ 
\hline
MA (id) & 10.7 (6.9) & 10.9 (4.0) & 12.0 (2.7) & 11.0 (7.1) & 11.2 (4.3) & 14.2 (4.3) & 12.9 (8.4) & 12.9 (6.2) & 14.2 (4.3) \\ 
 MA (id, last 50 \%) & 10.7 (6.5) & 12.0 (4.6) & 11.9 (2.7) & 11.0 (6.9) & 12.5 (5.0) & 15.4 (6.1) & 13.9 (9.6) & 15.1 (7.8) & 15.4 (6.1) \\ 
  MA (id, last 30 \%) & 11.9 (7.5) & 14.1 (6.5) & 12.9 (3.6) & 12.3 (8.0) & 14.7 (7.2) & 17.6 (8.3) & 16.0 (11.6) & 18.4 (10.6) & 17.6 (8.3) \\ 
\hline
  MA (logit) & 9.9 (5.8) & 11.3 (3.9) & 11.5 (2.2) & 10.2 (6.1) & 11.6 (4.2) & 13.7 (4.0) & 12.4 (8.0) & 13.3 (6.1) & 13.7 (4.0) \\ 
  MA (logit, last 50 \%) & 11.1 (6.5) & 12.8 (5.0) & 12.0 (2.8) & 11.4 (6.9) & 13.2 (5.5) & 15.4 (6.1) & 14.2 (9.7) & 15.7 (8.2) & 15.4 (6.1) \\ 
  MA (logit, last 30 \%) & 12.4 (7.9) & 14.8 (6.8) & 13.1 (3.8) & 12.8 (8.4) & 15.4 (7.5) & 17.6 (8.3) & 16.3 (11.9) & 18.7 (10.8) & 17.6 (8.3) \\ 
\hline
  MA (asin) & 10.2 (6.3) & 11.0 (3.9) & 11.7 (2.5) & 10.5 (6.5) & 11.3 (4.2) & 14.0 (4.2) & 12.4 (8.1) & 13.0 (6.1) & 14.0 (4.2) \\ 
  MA (asin, last 50 \%) & 10.8 (6.4) & 12.4 (4.8) & 12.0 (2.7) & 11.1 (6.9) & 12.8 (5.2) & 15.4 (6.1) & 14.0 (9.7) & 15.3 (8.0) & 15.4 (6.1) \\ 
  MA (asin, last 30 \%) & 12.1 (7.7) & 14.4 (6.6) & 13.0 (3.7) & 12.5 (8.2) & 15.0 (7.4) & 17.6 (8.3) & 16.1 (11.7) & 18.5 (10.7) & 17.6 (8.3) \\ 
\hline
  linear (id) & 15.2 (9.2) & 11.8 (7.5) & 8.2 (3.7) & 15.4 (9.2) & 12.3 (7.6) & 12.7 (6.0) & 17.0 (9.7) & 14.7 (8.5) & 12.7 (6.0) \\ 
  linear (logit) & 13.0 (7.8) & 10.2 (6.4) & 7.8 (2.8) & 13.4 (8.0) & 10.9 (6.5) & 12.4 (5.7) & 15.6 (9.2) & 14.0 (7.8) & 12.4 (5.7) \\ 
  linear (asin) & 14.0 (8.5) & 10.9 (6.8) & 8.0 (3.3) & 14.3 (8.5) & 11.4 (7.0) & 12.6 (5.8) & 16.2 (9.3) & 14.2 (8.2) & 12.6 (5.8) \\ 
\hline
  RCS (id)& 17.4 (9.0) & 13.7 (7.1) & 8.3 (4.3) & 17.8 (9.0) & 14.2 (7.3) & 14.1 (6.2) & 20.1 (9.7) & 17.5 (8.4) & 14.1 (6.2) \\ 
  RCS (logit) & 15.7 (8.4) & 12.6 (6.4) & 7.7 (3.8) & 16.1 (8.5) & 13.2 (6.6) & 13.9 (6.1) & 18.7 (9.4) & 16.8 (8.0) & 13.9 (6.1) \\ 
  RCS (asin) & 16.5 (8.6) & 13.0 (6.8) & 7.9 (4.0) & 16.9 (8.7) & 13.6 (6.9) & 14.0 (6.2) & 19.4 (9.5) & 17.1 (8.2) & 14.0 (6.2) \\ 
\hline
  FP2 (id) & 17.2 (8.8) & 13.3 (6.7) & 7.7 (3.6) & 17.5 (8.9) & 13.9 (6.9) & 13.4 (5.9) & 19.8 (9.7) & 17.1 (8.2) & 13.4 (5.9) \\ 
  FP2 (logit) & 15.4 (8.1) & 12.3 (6.1) & 7.6 (3.6) & 15.8 (8.2) & 12.9 (6.3) & 13.4 (5.9) & 18.4 (9.2) & 16.5 (7.8) & 13.4 (5.9) \\ 
  FP2 (asin) & 16.2 (8.4) & 12.7 (6.4) & 7.6 (3.6) & 16.6 (8.5) & 13.3 (6.6) & 13.4 (5.9) & 19.1 (9.4) & 16.8 (8.0) & 13.4 (5.9) \\ 
\end{tabular}}
\end{center}
\label{tab:areaK15}
\end{table}

\newpage

\begin{table}[!ht]
\caption{Results of the simulation study ($K=50$). The table summarizes the areas enclosed by the true and the estimated $C$-index curves, as obtained from the meta-regression models described in Section 2 of the paper. All areas were divided by the interval lengths $(\max_k (\tau_k)-\min_k (\tau_k))$ and multiplied by 1000.}
\begin{center}
\resizebox{\textwidth}{!}{
\begin{tabular}{l|rrr|rrr|rrr}
    \hline
    &\multicolumn{3}{c}{$\sigma_a=0$}&\multicolumn{3}{c}{$\sigma_a=0.01$}&\multicolumn{3}{c}{$\sigma_a=0.03$}\\
        \hline
     &short&moderate&long & short&moderate&long &short&moderate&long \\ \hline
    MA (id) & 8.0 (3.8) & 9.6 (1.3) & 12.0 (1.0) & 8.1 (3.8) & 9.7 (1.3) & 13.1 (1.8) & 8.7 (4.3) & 10.2 (2.1) & 13.1 (1.8) \\ 
  MA (id, last 50 \%) & 7.2 (2.8) & 11.5 (2.4) & 11.6 (1.0) & 7.4 (3.1) & 11.6 (2.6) & 13.5 (3.4) & 8.9 (4.7) & 12.6 (4.2) & 13.5 (3.4) \\ 
  MA (id, last 30 \%) & 8.2 (3.6) & 13.0 (3.2) & 12.1 (1.6) & 8.7 (4.1) & 13.3 (3.7) & 14.7 (4.7) & 10.8 (6.3) & 14.7 (5.9) & 14.7 (4.7) \\ 
\hline
  MA (logit) & 7.0 (2.5) & 10.6 (1.8) & 11.5 (0.7) & 7.2 (2.7) & 10.6 (1.9) & 12.5 (1.5) & 8.3 (3.9) & 10.8 (2.6) & 12.5 (1.5) \\ 
  MA (logit, last 50 \%) & 8.3 (3.6) & 12.5 (2.7) & 11.7 (1.2) & 8.7 (3.9) & 12.7 (3.1) & 13.5 (3.4) & 10.1 (5.6) & 13.3 (4.7) & 13.5 (3.4) \\ 
  MA (logit, last 30 \%) & 9.5 (4.4) & 13.9 (3.5) & 12.4 (1.8) & 10.0 (4.9) & 14.3 (4.0) & 14.7 (4.7) & 12.0 (6.9) & 15.3 (6.2) & 14.7 (4.7) \\ 
\hline
  MA (asin) & 7.1 (2.9) & 9.9 (1.4) & 11.7 (0.8) & 7.2 (3.0) & 9.9 (1.5) & 12.8 (1.7) & 8.0 (3.7) & 10.2 (2.1) & 12.8 (1.7) \\ 
  MA (asin, last 50 \%) & 7.6 (3.1) & 12.0 (2.6) & 11.6 (1.1) & 7.9 (3.4) & 12.1 (2.9) & 13.5 (3.4) & 9.3 (5.0) & 12.9 (4.4) & 13.5 (3.4) \\ 
  MA (asin, last 30 \%) & 8.8 (4.0) & 13.5 (3.3) & 12.2 (1.7) & 9.2 (4.4) & 13.8 (3.8) & 14.7 (4.7) & 11.2 (6.5) & 14.9 (6.0) & 14.7 (4.7) \\ 
\hline
  linear (id) & 12.8 (6.4) & 9.5 (5.3) & 6.5 (1.3) & 12.5 (6.4) & 9.5 (5.3) & 8.5 (2.8) & 11.8 (6.4) & 9.8 (5.4) & 8.5 (2.8) \\ 
  linear (logit) & 7.4 (4.5) & 5.7 (3.4) & 6.6 (1.0) & 7.6 (4.5) & 6.0 (3.5) & 8.3 (2.6) & 8.7 (5.1) & 7.9 (4.3) & 8.3 (2.6) \\ 
  linear (asin) & 9.9 (5.6) & 7.2 (4.4) & 6.4 (1.0) & 9.8 (5.6) & 7.4 (4.4) & 8.3 (2.6) & 9.9 (5.6) & 8.6 (4.7) & 8.3 (2.6) \\ 
\hline
  RCS (id) & 14.4 (5.6) & 11.4 (4.9) & 6.1 (2.6) & 14.4 (5.6) & 11.5 (4.9) & 9.5 (3.6) & 14.7 (5.7) & 12.6 (5.1) & 9.5 (3.6) \\ 
  RCS (logit) & 10.6 (4.5) & 8.5 (3.7) & 4.9 (2.0) & 10.8 (4.5) & 8.9 (3.8) & 9.0 (3.5) & 12.4 (5.0) & 11.1 (4.4) & 9.0 (3.5) \\ 
  RCS (asin) & 12.3 (5.1) & 9.7 (4.3) & 5.4 (2.2) & 12.4 (5.1) & 9.9 (4.3) & 9.2 (3.5) & 13.3 (5.3) & 11.7 (4.7) & 9.2 (3.5) \\ 
\hline
  FP2 (id)& 13.5 (5.8) & 10.5 (4.9) & 5.7 (2.3) & 13.3 (5.8) & 10.5 (4.9) & 8.3 (3.3) & 13.2 (5.9) & 11.2 (5.0) & 8.3 (3.3) \\
  FP2 (logit) & 9.1 (4.5) & 7.4 (3.6) & 4.8 (1.9) & 9.3 (4.5) & 7.7 (3.7) & 8.0 (3.3) & 10.7 (5.0) & 9.6 (4.3) & 8.0 (3.3) \\ 
  FP2 (asin) & 11.1 (5.2) & 8.6 (4.3) & 5.1 (2.0) & 11.0 (5.2) & 8.8 (4.3) & 8.1 (3.3) & 11.6 (5.4) & 10.2 (4.5) & 8.1 (3.3) \\ 
\hline

\end{tabular}}
\end{center}
\label{tab:areaK50}
\end{table}

\end{document}